\documentclass[12pt,letterpaper]{article}
\pdfoutput=1

\usepackage{graphicx,array}
\usepackage{color}
\usepackage{latexsym}
\usepackage{amsthm}
\usepackage{amsmath}
\usepackage{amssymb}
\usepackage[numbers,sort&compress]{natbib}
\usepackage{bm}
\usepackage{slashed} 
\usepackage{mathrsfs}
\usepackage{hyperref} 
\hypersetup{
    colorlinks=true,       
    linkcolor=red,          
    citecolor=blue,        
    filecolor=magenta,      
    urlcolor=blue           
}
\usepackage[all]{hypcap} 

\usepackage{myterms}
\newcommand{\CP}{\text{\sc cp}}
\newcommand{\xCP}{\slashed{\CP}}

\newcommand{\be}{\begin{equation}}
\newcommand{\ee}{\end{equation}}

\newcommand{\LB}{\text{\rm\tiny{L$\to$B}}}
\newcommand{\NL}{\text{\it\tiny\!{N$\rightarrow$L}}}
\newcommand{\LL}{\text{\it\tiny\!{${L}{\leftrightarrow}{\bar{L}}$}}}
\newcommand{\NN}{\text{\it\tiny\!{${N}{\leftrightarrow}{\bar{N}}$}}}

\newcommand{\LHN}{\text{\it\tiny\!\!{LHN}}}

\newcommand{\N}{\text{\it\tiny\!{N}}}

\newcommand{\wo}{\! {\rm w.o.}}

\newcommand{\aaa}{\Delta a}
\newcommand{\LLL}{\Delta L}
\newcommand{\HHH}{\Delta H}
\newcommand{\NNN}{\Delta N}
\newcommand{\EEE}{\Delta E}
\newcommand{\UUU}{\Delta U}
\newcommand{\DDD}{\Delta D}
\newcommand{\QQQ}{\Delta Q}

\title{\bf Baryogenesis at a Lepton-Number-Breaking Phase Transition}

\author{\large Andrew J. Long$^{a}$, Andrea Tesi$^{b}$, Lian-Tao Wang$^{a,b}$}
\date{\small \it 
$^a$Kavli Institute for Cosmological Physics, University of Chicago, Chicago, Illinois 60637, USA \\
$^b$Enrico Fermi Institute, University of Chicago, Chicago, Illinois 60637, USA
}

\begin{document}

\maketitle

\setlength{\parskip}{0.2ex}

\begin{abstract}
We study a scenario in which the baryon asymmetry of the universe arises from a cosmological phase transition where lepton-number is spontaneously broken.  
If the phase transition is first order, a lepton-number asymmetry can arise at the bubble wall, through dynamics similar to electroweak baryogenesis, but involving right-handed neutrinos.  
In addition to the usual neutrinoless double beta decay in nuclear experiments, the model may be probed through a variety of ``baryogenesis by-products'', which include a stochastic background of gravitational waves created by the colliding bubbles.
Depending on the model, other aspects may include a network of topological defects that produce their own gravitational waves, additional contribution to dark radiation, and a light pseudo-Goldstone boson (majoron) as dark matter candidate.
\end{abstract}

\newpage
\section{Introduction}\label{sec:Introduction}

Thermal leptogenesis has been a remarkably successful framework for explaining the origin of the matter / antimatter asymmetry of the early universe.  
In this scenario \cite{Fukugita:1986hr} a lepton asymmetry arises from the out of equilibrium and CP-violating decay of heavy, Majorana neutrinos, and it is processed into a baryon asymmetry by the electroweak sphaleron.  
In part, thermal leptogenesis is appealing because it requires only a minimal and well-motivated extension of the Standard Model (SM).  
Namely, the heavy Majorana neutrino fits naturally into the seesaw mechanism for explaining the mass scale of the light neutrinos.  
In this article we suppose that the Majorana mass arises from the vacuum expectation value of a scalar field, which spontaneously breaks lepton-number.  
We illustrate how baryogenesis could occur during the $\U{1}_{\rm L}$-breaking phase transition.  

Specifically, we suppose that the $\U{1}_{\rm L}$-breaking phase transition is first order.  
In the symmetric phase (outside the bubbles) the right-handed neutrinos are massless, and in the broken phase (inside the bubbles) they acquire a large Majorana mass.  
This leads to a CP-violating scattering of neutrinos from the expanding bubble wall, which generates a lepton asymmetry in front of the wall.  
The lepton asymmetry is transferred from the right-handed neutrinos to the SM leptons through the Yukawa interactions, and finally the lepton asymmetry diffuses into the bubble where it is eventually converted into a baryon asymmetry through the electroweak sphaleron.  
Although lepton-number is broken inside the bubble, washout is avoided because the phase transition is strongly first order, which means that the Majorana mass of the right-handed neutrinos, $m_N$, satisfies $m_N / T \gtrsim 10$.

In principle the out-of-equilibrium decay of the heavy right-handed Majorana neutrinos (inside the bubbles) can also contribute to the lepton asymmetry, just like in thermal leptogenesis.  
However, since the CP-violating decay of the lightest right-handed neutrino requires a loop containing one of the heavier right-handed neutrinos, the spectrum cannot be too hierarchical, otherwise the CP-violation parameter is suppressed by the small mass ratio.  
For additional details, see the review \cite{Strumia:2006qk}.  
In our model, CP-violation at the bubble wall receives no such suppression, and therefore we focus on the lightest right-handed neutrino and neglect an additional contribution to baryogenesis from its decay.  

The model we consider here shares common elements with an early implementation of electroweak baryogenesis by Cohen, Kaplan, \& Nelson (1990) \cite{Cohen:1990it, Cohen:1990py}.  
Both models generate a lepton asymmetry by the CP-violating scattering of right-handed neutrinos from the bubble wall.  
Whereas our model naturally operates at the seesaw scale, the model of Cohen et. al. operates at the weak scale where both the SM Higgs and a new $\U{1}_{\rm L}$-breaking scalar participate in the first order phase transition.  
However we notice that it is difficult to generate the known baryon asymmetry in the model of Refs.~\cite{Cohen:1990it, Cohen:1990py} while also inducing the light neutrino masses through the seesaw mechanism. A recent study \cite{Cline:2017qpe} drops the connection with neutrino physics by replacing the right-handed neutrino with a dark matter candidate.  
Another recent paper \cite{Pascoli:2016gkf} studies leptogenesis from a first-order phase transition in one of the complex phases of the dimension-5 Weinberg operator.

In the endeavor to solve the problem of baryogenesis, perhaps the greatest challenge is testability.  
It is desirable to have a theoretically-compelling model that is also accessible to laboratory and cosmological probes.  
The foundation of thermal leptogenesis is a well-motivated model of particle physics whose predictions for the mass and properties of the light neutrinos can be tested in the laboratory.  
Electroweak baryogenesis necessarily requires new physics at the weak scale, which we continue to explore with high energy collider experiments, but perhaps more important is that the first order electroweak phase transition can also generate various cosmological relics in addition to the matter / antimatter asymmetry, such as a stochastic background of gravitational waves.  
For additional details, see Refs.~\cite{Morrissey:2012db,Caprini:2015zlo}.  
Whereas laboratory measurements provide only indirect and model-dependent information about the conditions of the early universe, the observation of these ``baryogenesis by-products'' would provide a new, direct probe into the epoch of baryogenesis.  
Our model acquires a connection with neutrino physics (laboratory probes) through its common features with thermal leptogenesis, and we have a connection with baryogenesis by-products (cosmological probes) from similarities to electroweak baryogenesis

The organization of this paper is as follows.  
We discuss the baryogenesis mechanism in \sref{sec:Baryogen} in the context of a simplified toy model, and we estimate the predicted baryon asymmetry of the universe. 
In \sref{sec:Model} we discuss a concrete particle physics model in which the baryogenesis mechanism could be implemented.  
We highlight a few interesting aspects of the particle physics phenomenology and cosmology in \sref{sec:Pheno}.  
We close the article in \sref{sec:Discussion} with a brief summary and discussion of directions for future work.  


\section{Baryogenesis at a $\U{1}_{\rm L}$-breaking Phase Transition}\label{sec:Baryogen}

In this section we present the key components of our proposed baryogenesis mechanism without fully specifying the particle physics model.  
We flesh out the model-dependent details in Section~\ref{sec:Model}.  

\subsection{Overview of the mechanism}\label{sub:Overview}

We let the SM be extended to include a Weyl spinor field $N$ and a complex scalar field $S$, which are singlets under the SM gauge group.  
They have the following interactions 
\begin{align}\label{eq:L_int}
	- \mathscr{L}_{\rm int} = \frac{1}{2} \kappa S NN + \lambda_N L H N + \lambda_E L H^{\ast} E + \hc 
\end{align}
where $L_i$ and $E_i$ are the SM lepton doublet and singlet of generation $i$, and $H$ is the Higgs doublet.  
Here and in the following all the fermions are represented by left-handed, two-component Weyl spinors.
This lagrangian respects a $\U{1}_{\rm L}$ lepton number, under which the charge assignments are ${\rm L}(N) = -1$, ${\rm L}(S) = +2$, ${\rm L}(L_i) = +1$, ${\rm L}(E_i) = -1$, and ${\rm L}(H) = 0$.  
We assume the $\U{1}_{\rm L}$ symmetry to be spontaneously broken by the condensation of the scalar $S$.  
For the moment we do not specify the structure of the scalar potential, and we postpone this discussion to \sref{sec:Model}.  
After the spontaneous symmetry breaking the neutrino $N$ will get a Majorana mass $m_N= \kappa \langle S \rangle$, which is assumed to be above the weak scale $v \simeq 246 \GeV$.  
The light neutrino masses arise at low energy via the Type-I seesaw mechanism \cite{Minkowski:1977sc, Mohapatra:1979ia, GellMann:1980vs, Yanagida:1980xy, Mohapatra:1980yp, Schechter:1980gr}, and the coupling $\lambda_{N}$ is expressed as 
\begin{align}\label{eq:lamN}
	\lambda_N \approx \sqrt{ \frac{2m_N m_{\nu}}{v^2} } \simeq \bigl( 6 \times 10^{-2} \bigr) \sqrt{ \frac{m_N}{10^{12} \GeV} \frac{m_{\nu}}{0.1 \eV}} 
	\per 
\end{align}
where $m_{\nu} \simeq 0.1 \eV$ is the observed neutrino mass scale.  

\begin{figure}[t]
\begin{center}
\includegraphics[width=0.49\textwidth]{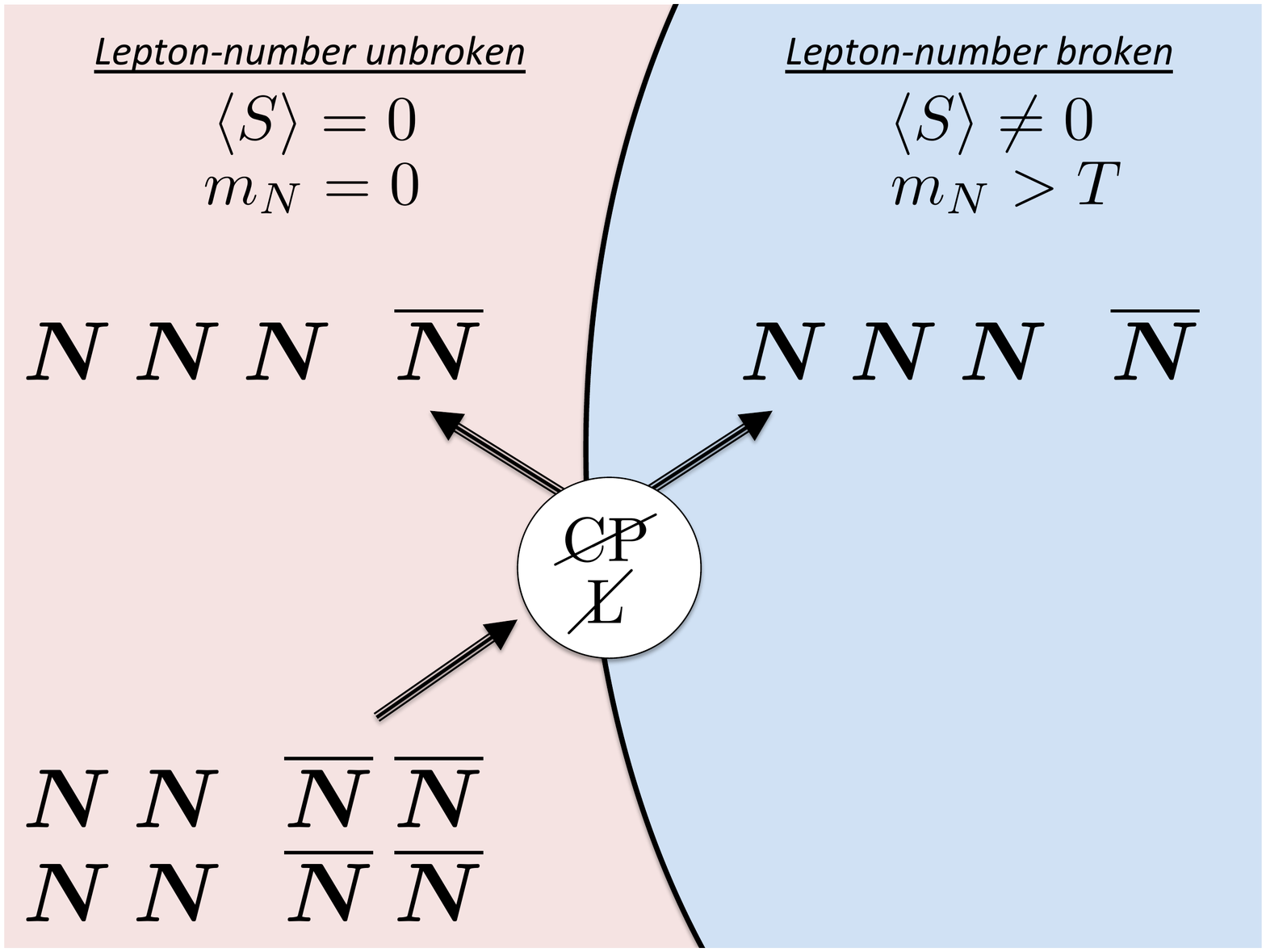} \hfill
\includegraphics[width=0.49\textwidth]{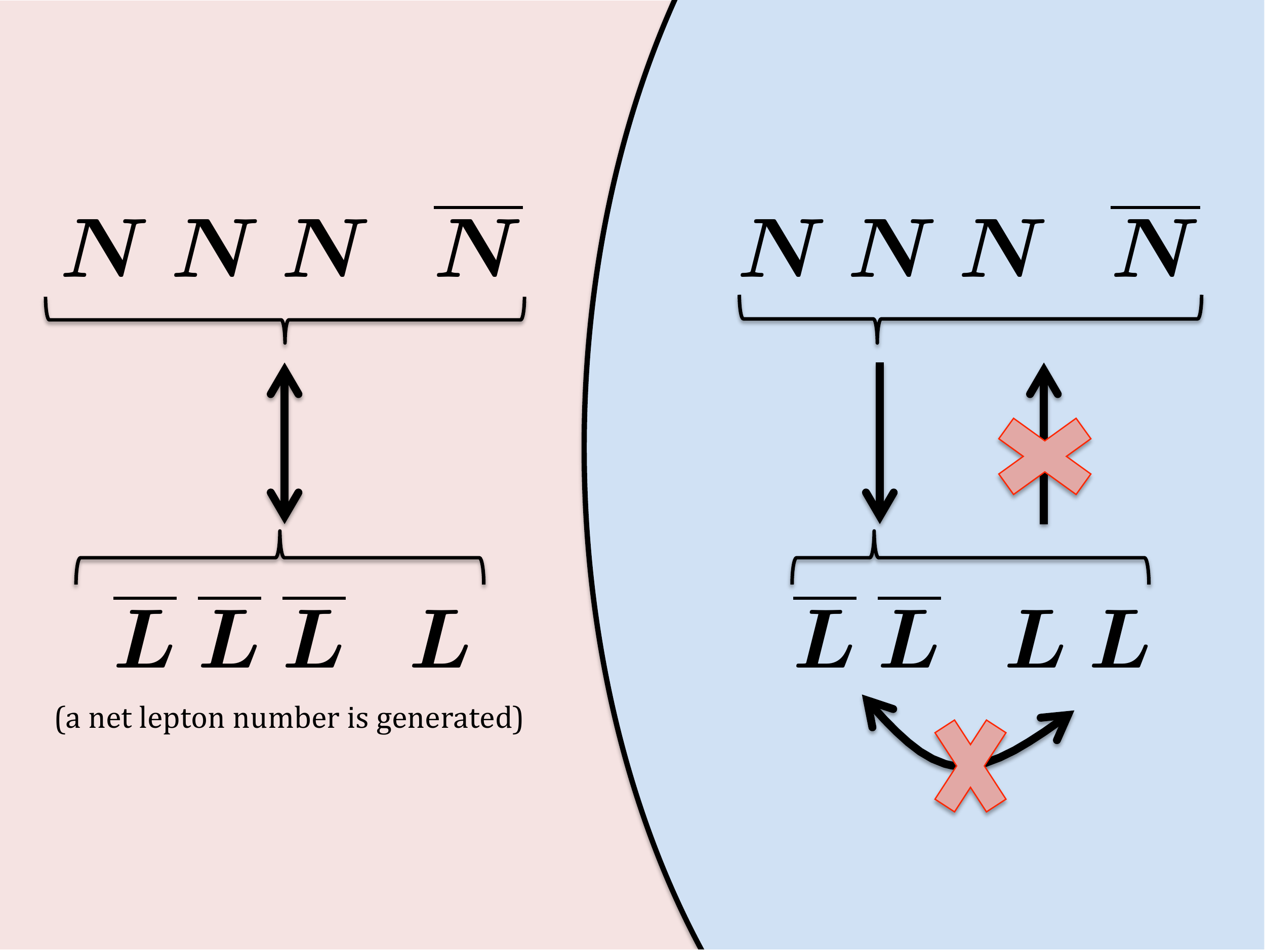}
\caption{\label{fig:cartoon}
This cartoon illustrates the stages of the leptogenesis mechanism discussed in the text.  {\it Left:}  The CP-violating scattering of right-handed neutrinos $\bar{N}$ and left-handed anti-neutrinos $N$ from the bubble wall generates a lepton-number.  {\it Right:}  Outside of the bubble, the lepton-number is transferred from $N$ to the SM left-handed leptons $L_i$ via lepton-number-preserving interactions.  Inside of the bubble, lepton-number-violating interactions are out of equilibrium, and the lepton asymmetry is not washed out.  
}
\end{center}
\end{figure}

The hot and dense conditions of the early universe caused the $\U{1}_{\rm L}$ to be restored.  
In the $m_N = 0$ phase the field $N$ describes two particles: a massless left-handed anti-lepton $N$ and a massless right-handed lepton $\bar{N}$.  
Initially $N$ and $\bar{N}$ are in thermal equilibrium at temperature $T$ with equal abundances.  
As the universe expanded and cooled, the $\U{1}_{\rm L}$ symmetry became spontaneously broken through a first order phase transition at temperature $T_{\rm L}$.  
Bubbles of the $m_N \neq 0$ phase nucleated in a background of the $m_N = 0$ phase, and they grew until they filled all of space and the phase transition was completed.  
During the phase transition, $N$ and $\bar{N}$ scatter from the bubble wall as illustrated in \fref{fig:cartoon}.  
If the interactions at the wall are CP-violating, the $\bar{N}$ are preferentially transmitted through the wall and the $N$ are preferentially reflected.  
Effectively, the wall sources $N$-number at a rate per unit volume that we denote by $S_{\N}^{\xCP}$.  
If the wall has thickness $L_w$ and moves with speed $v_w$ (in the rest frame of the plasma) then the volume of space occupied by the wall is exposed to the source for a time $L_w / v_w$.  

The sourced $N$-number diffuses away from the bubble wall.  
If the diffusion length is large, then some of the $N$-number will enter the bubble where it can be partially erased by $N$-number-violating interactions, which arise from the nonzero Majorana mass $m_N$.  
Consequently, the $N$-number density is suppressed by a factor that we denote as $\varepsilon_{\NN}$.  
In front of the wall, reactions mediated by the Yukawa interactions ($L H N$ and $L H^{\ast} E$) are active, and they transfer a fraction $f_{\NL}$ of the $N$-excess into the SM leptons.  
Behind the wall, the $\U{1}_{\rm L}$ symmetry is broken, and lepton-number-violating scatterings such as $L_i H \leftrightarrow \bar{L}_j \bar{H}$ threaten to wash out the lepton asymmetry.  
In general, washout suppresses the lepton number by a factor of $\varepsilon_{\LL}$.  
Provided that the Majorana mass is sufficiently large inside the bubbles, $m_N \gg T $, washout is avoided and the phase transition is said to be ``strongly'' first order.  

Finally a fraction $f_{\LB}$ of the lepton asymmetry is converted into a baryon asymmetry by the electroweak sphalerons.  
The resulting baryon-to-entropy ratio can be written schematically as 
\begin{align}\label{eq:nB_ov_s}
	\frac{n_{\rm B}}{s} = f_{\LB} \, \varepsilon_{\LL} \, f_{\NL} \, \varepsilon_{\NN} \, \frac{L_w}{v_w} \, \frac{S_{\N}^{\xCP}}{s} 
\end{align}
where $n_{\rm B}$ is the number density of baryon number and $s$ is the entropy density of the plasma after the $\U{1}_{\rm L}$ phase transition is complete.  
Note there is additional dependence on $v_w$ and $L_w$ in the various factors, and the scaling with these parameters is not obvious from \eref{eq:nB_ov_s}.  
In the following subsections we estimate each of these factors.  

\subsection{CP-Violating Phase Gradient}\label{sub:CPV_Phase}

In order for the scattering of $N$ and $\bar{N}$ from the bubble wall to violate CP, it is necessary that the Majorana mass of $N$ has a nontrivial phase gradient.  
In this section we discuss how the phase gradient arises, and in the next section we discuss how it leads to CP-violating scattering.  

During the first order $\U{1}_{\rm L}$-breaking phase transition, the scalar field expectation value becomes inhomogeneous $\langle S(x)\rangle = v_S(x) e^{i \theta(x)} / \sqrt{2}$.  
Through the Yukawa interaction in \eref{eq:L_int}, this leads to an inhomogeneous Majorana mass $m_N(x) e^{i \theta(x)}$ where $m_N(x) = \kappa v_S(x) / \sqrt{2}$ is real.  
In the phase of unbroken $\U{1}_{\rm L}$ we have $m_N(x) = 0$ and $\theta(x) = 0$, and at the interface with the phase of broken $\U{1}_{\rm L}$, {\it i.e.} the bubble wall, the profile functions rise smoothly, eventually reaching their temperature-dependent asymptotic values $m_N(T)$ and $\theta(T)$ inside the bubble.  

On scales that are small compared to the curvature of the bubble, we can treat the bubble wall as planar.  
Without further loss of generality we can move to a frame where the wall is at rest and oriented normal to the $z = x^3$ axis.  
Let the wall thickness be denoted by $L_w$.  
For the sake of discussion, we will demarcate $z < -L_w/2$ as the phase of unbroken $\U{1}_{\rm L}$ (in front of the wall, outside of the bubble) and $L_w/2 < z$ as the phase of broken $\U{1}_{\rm L}$ (behind the wall, inside of the bubble).  
In the rest frame of the plasma, the wall moves with speed $v_w$ into the phase of unbroken $\U{1}_{\rm L}$.  

We can describe the interaction of $N$ with the wall using the low energy effective theory.  
After a (coordinate-dependent) rephasing\footnote{
The rephasing affects also the SM leptons, and they would feel a CP violating background that modifies their dispersion relations, however they have completely negligible interactions with the bubble wall.  }
 $N \to N \, e^{-i \theta(x)/2}$, the effective theory for the $N$ in this background is
\begin{align}\label{eq:L_eff}
	\mathscr{L}_{\rm eff} \supset i N^{\dagger} \bar{\sigma}^{\mu} \partial_{\mu} N + \frac{1}{2} \partial_{\mu} \theta j_N^{\mu} -  \frac{1}{2} m_N(x) \, (NN + \hc ) 
\end{align}
where $j_N^{\mu} = N^{\dagger} \bar{\sigma}^{\mu} N$ is the $N$-number current density.  
Since $j_{N}^{\mu}$ is a chiral current, the coordinate-dependent profile for $\theta$ breaks CP.  
If $\theta$ were homogeneous and/or the current exactly conserved, then we would get no physical effect.  

The physical CP-violating effect of the phase gradient is captured by the dispersion relation.  
Using \eref{eq:L_eff} the kinetic term can be written as $i N^{\dag}\bar{\sigma}^{\mu} (\partial_{\mu} - i \partial_{\mu} (\theta/2)) N$.  
In the rest frame of the wall, we can write $\partial_\mu \theta = (0,0,0,\theta^{\prime})$ where $\theta^{\prime} = d\theta/dz$.  
Parametrically, $\theta^{\prime} \approx \theta(T) / L_w$ at the wall, and $\theta^{\prime} \approx 0$ either inside or outside of the bubble.  
The spatial gradient affects the propagation of the spin up and down components differently, since it splits the energy $E\to E \pm \theta^{\prime}/2$.  
If the bubble wall is viewed as a potential energy barrier, then the phase gradient lowers the height of the barrier for one helicity and raises it for the other.  


\subsection{CP-Violating Scattering and Source of $N$-Number}\label{sub:CPV_Source}

The CP-violating phase gradient allows $\bar{N}$ to be converted into $N$ at the bubble wall.  
Effectively the wall acts as a source of $N$-number, {\it i.e.} the quantum number that counts $+1$ for $N$ and $-1$ for $\bar{N}$.  
In this section we calculate that source, denoted $S_{\N}^{\xCP}$. 

A simplified description of what is happening at the boundary between the two pahses consists in considering an $\bar{N}$ that is incident on the wall.  
This particle can either pass through the wall remaining an $\bar{N}$, or it can experience a $\Delta {\rm L} = -2$ interaction with the wall-forming fields and be reflected back as an $N$.  
We denote the ``reflection'' probability by $\Rcal$, and we let $\bar{\Rcal}$ denote the probability for an incident $N$ to reflect as an $\bar{N}$.  
If the scattering respects CP then $\Rcal = \bar{\Rcal}$, and the fluxes of $N$ and $\bar{N}$ from the wall are equal.  
However, a CP-violating phase gradient allows $\Rcal \neq \bar{\Rcal}$.  

We calculate the $N$-number source following the formalism of Refs.~\cite{Huet:1995sh,Huet:1995mm}, but see also Refs.~\cite{Riotto:1997vy,Lee:2004we} for a different approach using the closed time path (CTP) formalism.  
The CP-violating source is given by a thermal average of the differential reflection probability $\Rcal - \bar{\Rcal}$.  
To perform the thermal averaging, we require the phase space distribution function of $N$ in the rest frame of the wall.  
In the rest frame of the {\it plasma}, the distribution functions take the Fermi-Dirac form with temperature $T$, and boosting with a speed $v_w$ in the $-z$ direction gives the distribution function in the rest frame of the wall\footnote{To a good approximation, the distributions of $N$ and $\bar{N}$ are identical, {\it i.e.} $f_{\bar{N}} \approx f_N$.  The asymmetry being generated at the wall is assumed to be negligible.  }
\begin{align}\label{eq:fN_wall}
	f_N(z,p_x,p_y,p_z) &= \Bigl[ \mathrm{exp}\big(\gamma_w (E(z) - v_w p_{z} ) / T \big) + 1 \Bigr]^{-1} 
\end{align}
where $\gamma_w = 1 / \sqrt{1 - v_w^2}$ is the boost factor, and $E(z) = \sqrt{ |{\bf p}|^2 + m_N(z)^2 }$.  
Due to the extra term, $-v_w p_z$, particles with $p_z \gtrsim 0$ are slightly more abundant than particles with $p_z \lesssim 0$.  
In other words, there are more particles incident on the wall from outside the bubble than from inside the bubble.  
This is perhaps easier to understand in the rest frame of the plasma where the wall moves with speed $v_w$, and because the particle velocities follow an approximate Boltzmann distribution, there are more particles with speed $v < v_w$ than $v > v_w$.  

The source term can be written as \cite{Huet:1995sh,Huet:1995mm} 
\begin{align}\label{eq:SCP_def}
	S_{\N}^{\xCP}(z) = \frac{2}{\tau} \int_{-\infty}^{\infty} \frac{\ud p_x}{2\pi} \int_{-\infty}^{\infty} \frac{\ud p_y}{2\pi} \int_{0}^{\infty} \frac{\ud p_z}{2\pi} 
	\, \delta \! f_N
	\bigl( \Rcal - \bar{\Rcal} \bigr) 
	\per
\end{align}
where $\delta \! f_N\equiv f_N(z,p_x,p_y,p_z) -  f_N(z + \Delta,p_x,p_y,-p_z)$ accounts for the variation of the thermal distribution in a section of the bubble wall of length $\Delta$, where we assume $\Delta \ll L_w$.  This formula takes into account contribution from particles crossing the section from both directions \cite{Huet:1995sh,Huet:1995mm}. 

The length $\Delta$ is physically related to the mean free path of $N$ particles in the thermal bath and it is set by the rate of incoherent scatterings with the plasma.  
An important time scale is set by the thermalization time scale denoted by $\tau$, that relates $\Delta(z) = \tau v_z$ where $v_z = p_z / E(z)$ is the component of velocity normal to the wall.  
Thermalization primarily occurs through scatterings such as $NS \leftrightarrow \bar{N}$, $NL \leftrightarrow \bar{H}$, and $NH \leftrightarrow \bar{L}$; thus we estimate $\tau^{-1} \sim {\rm max}\bigl[ \kappa^2 \, , \, \lambda_N^2 \bigr] T / 4\pi $. 

To evaluate the differential reflection probability $\Rcal - \bar{\Rcal}$ we work in the thick wall regime where $\Delta \ll L_w$.  
In this case, the probability can be estimated as $\Rcal = \big| \int_z^{z+\Delta} e^{-i 2p_z z'} m_N(z') e^{i\theta(z')} dz' \big|^2$, and $\bar{\Rcal}$ is obtained by sending $i \theta(z) \to - i \theta(z)$ \cite{Huet:1995sh,Huet:1995mm}.  
The differential reflection probability $\Rcal - \bar{\Rcal}$ is estimated to be
\begin{align}\label{eq:R_Rbar}
	\Rcal-\bar{\Rcal} & \approx 
	2 \frac{g\bigl( p_z \Delta(z) \bigr)}{p_z^3} \, m_N(z)^2 \frac{d\theta}{dz} 
	\per
\end{align}
where $g(\xi) \equiv (\sin \xi - \xi \cos \xi) \sin \xi$. 
We evaluate $\Rcal - \bar{\Rcal}$ by treating the mass insertion perturbatively, and therefore \eref{eq:R_Rbar} becomes increasingly reliable in the regime $p_z \gg m_N(z)$.  
\eref{eq:R_Rbar} explicitly shows that the non-trivial CP interference comes from a $z$-dependent phase in the mass term.  

Now we evaluate the $N$-number source from \eref{eq:SCP_def}.  
We can simplify the factor of $\delta \! f_N$ by assuming that the wall motion (in the rest frame of the plasma) is non-relativistic, $v_w \ll 1$, and that the mass gradient is negligible, $m_N(z) \approx m_N(z+\Delta)$.   
Then using \eref{eq:R_Rbar} the source can be written as 
\begin{align}
	S_{\N}^{\xCP} & \approx \frac{2}{\pi^2} \frac{\gamma_w v_w}{\tau} m_N(z)^2 \frac{d\theta}{dz} \, \Ical\bigl( m_N(z)/T \, , \, T \tau \bigr) 
\end{align}
where temperature dependence is captured by the integral 
\begin{align}\label{eq:I_def}
	\Ical(x,y) & \equiv 
	\int_{x}^{\infty} \! \frac{ \varepsilon \ud \varepsilon}{\sqrt{\varepsilon^2 - x^2}} \, \int_{0}^{1} \! \ud \cos \theta \, 
	\frac{\sqrt{1-\cos^2\theta}}{\cos^2 \theta}
	\frac{e^{\varepsilon}}{\bigl( e^{\varepsilon} + 1 \bigr)^2} \, 
	g\bigl( \frac{\bigl( \varepsilon^2 - x^2 \bigr) \cos^2 \theta}{\varepsilon} y \bigr)
	\per
\end{align}
We are unable to evaluate the integral in \eref{eq:I_def} analytically, but we have verified that the modulus of the integral is well-approximated by the empirical formula $\bigl| \Ical(x,y) \bigr| \sim \, {\rm min}\bigl[ y^4 \, , \, 0.1 y^{0} \bigr] (x \, e^{-x})$ in the parameter regime of interest.  
The source's $z$-dependent profile is controlled by $m_N(z)$, which goes to $0$ in front of the wall, and the phase gradient $d\theta/dz$, which only has support at the wall.  
In the next section, we will simplify by assuming that $S_{\N}^{\xCP}(z)$ has a top hat profile, which takes a constant value at the wall and vanishes elsewhere.  
The amplitude of the CP-violating $N$-number source at the wall is estimated as 
\begin{align}\label{eq:SCP_approx}
	S_{\N}^{\xCP}(T) & \approx 2 \frac{\gamma_w v_w}{\pi^2} \, m_N(T)^3 \, \frac{\theta(T)}{L_w} \, {\rm min}\bigl[ (T\tau)^3 \, , \, 0.1 (T\tau)^{-1} \bigr] \, e^{-m_N(T)/T} 
\end{align}
where we have estimated $d\theta/dz \approx \theta(T) / L_w$.  
Here we are being somewhat conservative by replacing $m_N(z)$ with its asymptotic value inside the bubble, $m_N(T)$.  
In this way, we underestimate the source through the exponential factor $e^{-m_N(T)/T} \leq e^{-m_N(z)/T}$.  
We expect that a more careful treatment, which retains the full $z$-dependent profile of the source, will lead to a larger final baryon asymmetry.  

\subsection{Lepton-Number Diffusion and Redistribution}\label{sub:Diffusion}

The sourced $N$-number diffuses in front of the bubble wall where it is partially transferred to the SM leptons, $L$ and $E$.  
This process is described by a system of transport equations.  
In this section we write down the transport equations, solve for the spatial distribution of $N$-number, and solve for the conversion into $L$-number in front of the wall.  
Let $n_N$ and ${\bm j}_N$ be the number density and current density of $N$-number in the rest frame of the plasma. 
In the diffusion approximation we can write ${\bm j}_N = - D_{N} \, {\bm \nabla} \, n_N$ where $D_{N}$ is the diffusion coefficient\footnote{The diffusion coefficient can be expressed as a combination of the root-mean-square velocity of the particle and its mean free path, $D = \lambda_{\rm mfp} v_{\rm rms}$.  The mean free path is inverse proportionally to the number density of scatterers and the total cross section, $\lambda_{\rm mfp} = 1 / ( \sigma_{\rm tot} n_{\rm scat})$.} for species $N$.  
The diffusion occurs through scatterings such as $NS \to NS$, $NL \to NL$, and $NH \to NH$, and therefore we estimate $D_{N}^{-1} \sim {\rm max}\bigl[ \kappa^4 \, , \, \lambda_N^4 \bigr] (4\pi)^{-2} T$.  

In the plasma frame, $n_N$ depends on the temporal coordinate $x^0$ and the spatial coordinate $x^3$ normal to the wall.  
However, in the rest frame of the wall, the density only depends on the spatial coordinate normal to the wall, denoted by $z$.  
Performing the appropriate Lorentz transformation, we can write $z = \gamma_w ( x^3 + v_w x^0)$ where $\gamma_w = 1 / \sqrt{1 - v_w^2}$ is the boost factor.  
Following the standard formulation, we write the transport equations in the rest frame of the plasma, but we express the $N$-number density in terms of $z$.  

The full system of transport equations are derived in \aref{app:transport}, and here we simply carry over the relevant results.  
The transport equation for $n_N$ encodes the various $N$-number-changing reactions in which $N$ and $\bar{N}$ participate.  
These include lepton-number-conserving reactions, such as $S \leftrightarrow NN$ and $H \leftrightarrow \bar{L}_i \bar{N}$, as well as lepton-number-violating reactions, such as $N \bar{N} \leftrightarrow NN$ and $H \leftrightarrow \bar{L}_i N$.  
We assume that the reactions with $S$ are fast and the reactions with $L$ are slow.  
Then the transport equation for $n_N$ is put into the simplified form
\begin{align}\label{eq:nN_eqn}
	v_{w} n_{N}^{\prime} - D_{N} n_{N}^{\prime \prime} 
	& \approx - \Gamma_{\N} \, n_N + S_{\N}^{\xCP} 
	\com
\end{align}
where $n_N$, $\Gamma_{\N}$, and $S_{\N}^{\xCP}$ are functions of the spatial coordinate $z$, and the prime denotes $d/dz$.  
The transport coefficient $\Gamma_{\N}(z)$ is the effective rate of $N$-number violation due to lepton-number-violating interactions behind the wall.  
The $N$-number source $S_{\N}^{\xCP}(z)$ was discussed in \sref{sub:CPV_Source}.  

A general solution of \eref{eq:nN_eqn} is available in \rref{Huet:1995sh}, and here we derive an approximate solution.  
The source $S_{\N}^{\xCP}(z)$ is localized at the wall, and therefore we approximate $S_{\N}^{\xCP}(z) = S_{\N}^{\xCP}(T)$, given by \eref{eq:SCP_approx}, for $-L_w/2<z<L_w/2$ and $S_{\N}^{\xCP}(z) = 0$ elsewhere.  
The washout term $\Gamma_{\N}(z)$ is active at the wall and inside the bubble, and therefore we approximate $\Gamma_{\N}(z) = \Gamma_{\N}(T)$ for $-L_w/2 < z$ and $\Gamma_{\N}(z) = 0$ elsewhere.  
We estimate $\Gamma_{\N}(T) \sim m_N(T)^2 / (10 T)$ \cite{Chung:2009qs, Konstandin:2013caa}.  
With these simplifications, it is straightforward to solve \eref{eq:nN_eqn} for $n_N(z)$.  
In front of the bubble wall, $z < -L_w/2$, the $N$-number density profile takes the form
\begin{align}\label{eq:nN_soln}
	n_N(z) \approx {\rm min}\big[1,\frac{1}{\sqrt{\Gamma_{\N} D_{N}/v_w^2}} \bigr] \, \frac{L_w}{v_w} \, S_{\N}^{\xCP} \, e^{v_w z / D_{N}} 
	\com
\end{align}
and we define $\varepsilon_{\NN} = {\rm min}\bigl[ 1 , \, 1 / \sqrt{\Gamma_{\N} D_{N} / v_w^2} \bigr]$.  
Due to the diffusion\footnote{The length scale $D_{N}/v_w$ and the time scale $D_{N}/v_w^2$ can be understood as follows.  In a time interval $\Delta t$ the wall moves a distance $\Delta z_{\rm wall} = v_w \Delta t$ and the sourced $N$-number diffuses a distance $\Delta z_{\rm diff} = \sqrt{ 2 D_{N} \Delta t }$ away from the wall.  Initially, $\Delta z_{\rm diff} > \Delta z_{\rm wall}$ but the wall catches up to the diffusing $N$-number after a time $\Delta t = \tau_{\rm diff}$ with $\tau_{\rm diff} \equiv D_{N} / v_w^2$ when it has moved a distance $\Delta z = L_{\rm diff}$ with $L_{\rm diff} \equiv D_{N} / v_w$.  }  the $N$-number precedes the wall for a distance $D_{N} / v_w$, which we have assumed to be much greater than the wall thickness $L_w$.  
The prefactor leads to a suppression of the $N$-number if the washout time scale $\Gamma_{\N}^{-1}$ is much shorter than the diffusion time scale $D_{N} / v_w^2$.  

In front of the wall, the $N$-number pushes reactions such as $N L_i \leftrightarrow \bar{H}$ and $N L_i \leftrightarrow \bar{H} W$ out of equilibrium.  
As these reactions re-equilibrate, the $N$-number excess is partially transferred to the SM lepton doublets $L_i$.  
To estimate the resultant $L$-number, we simplify the transport equations by focusing on the source term associated with the $N$-number excess.  
Let $n_L$ be the number density of $L$-number, which is summed over the $2$ isospin degrees of freedom and the $3$ generations.  
The simplified transport equation for $L$-number takes the form 
\begin{align}\label{eq:nL_eqn}
	v_{w} n_{L}^{\prime} - D_{L} n_{L}^{\prime \prime} \approx - \Gamma_{\LHN} \, n_{N} 
	\com
\end{align}
where $D_{L}^{-1} \sim \alpha_w^2 T$ is the lepton doublet diffusion coefficient \cite{Joyce:1994zt}, and $\Gamma_{\LHN}$ is the thermally averaged interaction rate.  
We evaluate $\Gamma_{\LHN}$ in \aref{app:transport} finding
\begin{align}\label{eq:Gam_LHN}
	\Gamma_{\LHN} = \frac{\lambda_N^2}{24\pi \zeta(3)} \frac{m_H(T)^3}{T^2} K_1\bigl( m_H(T) / T \bigr)
\end{align} 
where $K_n(x)$ is the modified Bessel function of the second kind and order $n$, and $m_H(T) \simeq 0.6 T$ is the thermal mass of the Higgs.  
We solve \eref{eq:nL_eqn} for $n_L$ in the background of the $N$-number density given by \eref{eq:nN_soln}.  
The $L$-number density at the bubble wall is found to be 
\be\label{eq:nL_soln}
	n_{L} \approx -  \mathrm{min}\Big[ 1,\,  \Gamma_{\LHN} \frac{D_{N}}{v_w^2}\Big]   \, n_{N}
	\com
\ee
and we define the $N$-to-$L$ conversion efficiency factor to be $f_{\NL} = - {\rm min}\bigl[ 1 \, , \, \Gamma_{\LHN} D_{N} / v_w^2 \bigr]$.  
Here we have taken the limits $L_w \ll D_{N}/v_w$, which is the case for the parameters of interest.  
In the regime $\Gamma_{\LHN} D_{N} / v_w^2 \gg 1$, the conversion is efficient, and an $O(1)$ fraction of the $N$-number will be converted to $L$-number.  

\subsection{Washout Avoidance}\label{sub:Washout}

Finally the lepton-number diffuses inside the bubble where $\U{1}_{\rm L}$ is broken.  
The scattering of SM leptons $L_i$ mediated by a Majorana neutrino $N$ threatens to washout the lepton-number.  
In this section, we estimate the washout factor and derive a condition on the Majorana mass to ensure that washout is avoided.  

Let $n_{\rm lep}=n_{L} - n_{E}$ denote the number density of SM lepton-number.  
At the bubble wall we have initially $n_{\rm lep} \approx n_{L}$ where $n_{L}$ is given by \eref{eq:nL_soln}.  
Inside the bubble, the evolution of $n_{\rm lep}$ is described by the kinetic equation
\begin{align}\label{eq:nlep_eqn}
	\dot{n}_{\rm lep} + 3 H n_{\rm lep} = - \Gamma_{\wo} n_{\rm lep} 
\end{align}
where $H(t)$ is the Hubble parameter at time $t$ and $\Gamma_{\wo}(t)$ is the thermally averaged rate of lepton-number-violating interactions.  
The solution of \eref{eq:nlep_eqn} is simply 
\begin{align}\label{eq:nlep_soln}
	n_{\rm lep}(t) = n_{\rm lep}(t_i) \, \left( \frac{a(t)}{a_i} \right)^{-3} {\rm exp} \Bigl[ - \int_{a_i}^{a(t)} \frac{\ud a^{\prime}}{a^{\prime}} \frac{\Gamma_{\rm w.o}(a^{\prime})}{H(a^{\prime})} \Bigr]
	\com 
\end{align}
where we have introduced the scale factor $a(t)$ and used $H(t) = \dot{a}/a$.  
At late times, the exponential factor becomes a constant, and we define the washout factor $\varepsilon_{\LL}$ to equal this constant.  
Assuming that the expansion of the universe is adiabatic, $\ud a/a = -\ud T/T$, we have\begin{align}\label{eq:epsLL}
	\varepsilon_{\LL} = {\rm exp} \Bigl[ - \int_{0}^{T_{\rm L}} \frac{\ud T}{T} \frac{\Gamma_{\rm w.o}(T)}{H(T)} \Bigr]
\end{align}
where $3 \Mpl^2 H(T)^2 = (\pi^2/30) g_{\ast} T^4$ with $\Mpl \simeq 2.43 \times 10^{18} \GeV$ and $g_{\ast} \simeq 106.75$.  

\begin{figure}[t]
\begin{center}
\includegraphics[height=8cm]{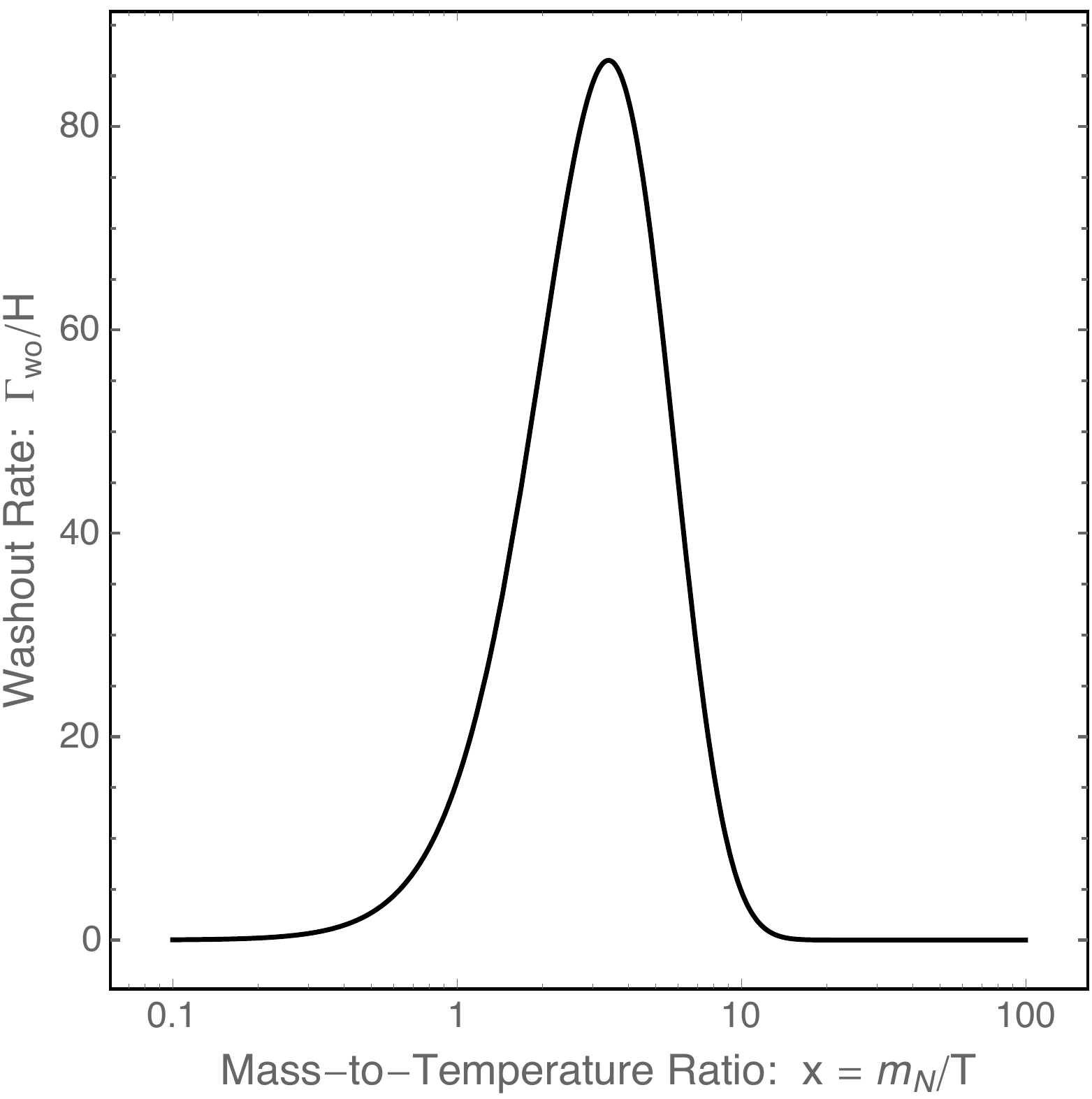} \hfill
\includegraphics[height=8cm]{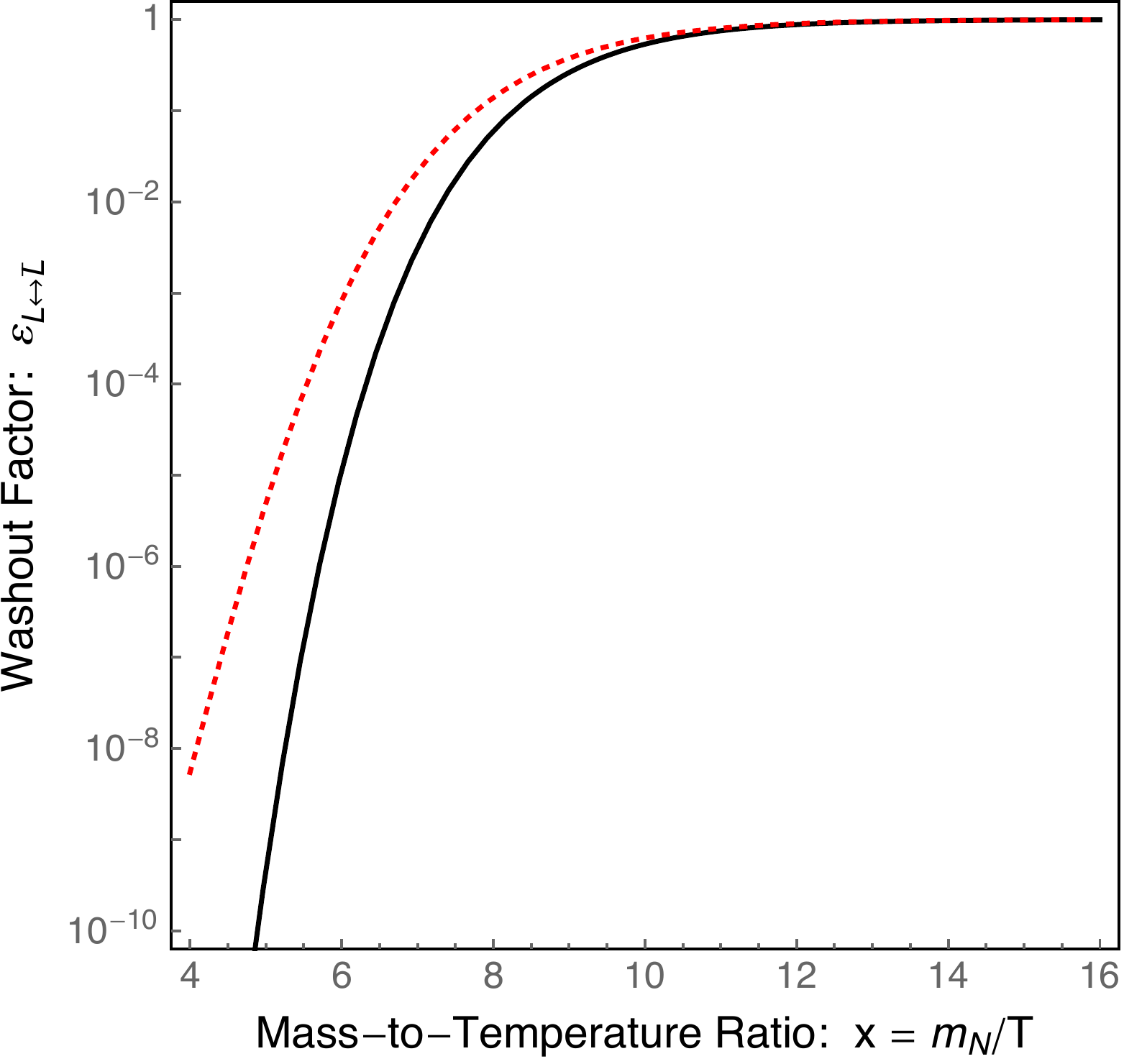}
\caption{\label{fig:washout}
The left panel shows the washout rate from \eref{eq:Gamma_wo_def} normalized to the Hubble parameter.  The right panel shows the washout factor from \eref{eq:epsLL} in black and the approximation from \eref{eq:epsLL_approx} in red (dashed).  The washout of lepton-number is avoided provided that the Majorana neutrino mass $m_N$ is sufficiently large inside of the bubbles.  
}
\end{center}
\end{figure}

The dominant contributions to lepton-number washout are the reactions $N \leftrightarrow L_i H$ and $\bar{N} \leftrightarrow \bar{L}_i \bar{H}$.  
We evaluate the thermally-averaged washout rate $\Gamma_{\wo}$ in \aref{app:transport}, and the result is found to be
\begin{align}\label{eq:Gamma_wo_def}
	\Gamma_{\wo} 
	= \frac{\lambda_N^2}{24\pi \zeta(3)} \frac{m_N(T)^3}{T^2} K_1\bigl( m_N(T) / T \bigr)
	\per
\end{align}
Using the expression for $\lambda_N$ that appears in \eref{eq:lamN}, we find $\Gamma_{\wo} / H \simeq 25.9 \, x^4 \, K_1(x)$ where $x = m_N(T)/T$.  
To avoid washout we need $\varepsilon_{\LL} \approx 1$, as shown in \fref{fig:washout}.  
The washout avoidance condition can be roughly expressed as $\Gamma_{\wo} < x H$, which implies $m_N(T_{\rm L}) / T_{\rm L} \gtrsim 9$ where $T_{\rm L}$ is the temperature of the $\U{1}_{\rm L}$ phase transition.  
This condition defines a ``strongly first order'' $\U{1}_{\rm L}$ phase transition.\footnote{For comparison, one usually defines a strongly first order {\it electroweak} phase transition by the requirement that electroweak sphaleron processes are out of equilibrium in the phase of broken electroweak symmetry.  This implies a lower bound on the sphaleron energy, $E_{\rm sph}(T) / T \gtrsim 40$ \cite{Bochkarev:1990fx}.}  
If the washout avoidance condition is satisfied, then the washout processes are out of equilibrium, and we can approximate \eref{eq:epsLL} as 
\begin{align}\label{eq:epsLL_approx}
	\varepsilon_{\LL} \approx {\rm exp}\Bigl[ - 32.5 \, \left( \frac{m_{\nu}}{0.1 \eV} \right) \left( \frac{g_{\ast}}{106.75} \right)^{-1/2} \, x^{5/2} \, e^{-x} \Bigr]_{x = m_N(T_{\rm L}) / T_{\rm L}} 
	\per
\end{align}
If the washout avoidance condition is not satisfied then the washout suppression factor $\varepsilon_{\LL}$ is given by \eref{eq:epsLL}.  

Let us briefly comment on the result in \eref{eq:epsLL_approx}.  
Note that the washout suppression factor is exponentially sensitive to the value of the lightest neutrino mass.  
This is because $\lambda_N^2 \propto m_{\nu}$ through the seesaw relation (\ref{eq:lamN}).  
It may be possible to alleviate the washout by lowering $m_{\nu}$, but this will also suppress the $L \to N$ conversion as seen in \eref{eq:Gam_LHN}, and we do not explore this limit in detail.  
Second, let us remark that washout leads to an exponential suppression (\ref{eq:nlep_soln}), because the washout processes remain active while the source is no longer present.  
This should be contrasted with the case of thermal leptogenesis (out of equilibrium, CP-violating Majorana neutrino decay) in which the source and washout processes are active simultaneously, and the suppression is only a power law for a large range of masses.  

\subsection{Relic Baryon Asymmetry}\label{sub:BaryonAsym}

Finally, the lepton-number that survives washout inside the bubbles is partially converted into baryon-number by the electroweak sphaleron \cite{Kuzmin:1985mm}.  
Let $n_{\rm lep}$ be the initial number density of lepton-number, and let $n_{\rm B}$ be the number density of baryon-number after the conversion.  
If only SM degrees of freedom are in equilibrium, then we have the relation \cite{Harvey:1990qw}
\begin{align}\label{eq:f_LB}
	n_{\rm B} = -\frac{28}{79} \, n_{\rm lep}
	\com 
\end{align}
and we define the conversion factor to be $f_{\LB} = -28/79$.  
This is the case for our scenario where the sphaleron conversion will continue after the lepton-number-breaking phase transition is completed and the new physics degrees of freedom have gone out of equilibrium.  
Notice also that the sphaleron transitions are in equilibrium starting from temperatures $T \lesssim 10^{10} \GeV$.

Drawing on the calculations in the previous sections, we estimate the relic baryon asymmetry as in \eref{eq:nB_ov_s} where the various factors appear in Eqs.~(\ref{eq:SCP_approx},~\ref{eq:nN_soln},~\ref{eq:nL_soln},~\ref{eq:epsLL},~\ref{eq:f_LB}).  
An approximate analytic solution for the baryon asymmetry is then given by
\begin{align}\label{eq:nB_final}
	\frac{n_{\rm B}}{s} & \approx 
	\pm \frac{28}{79} 
	\times {\rm min}\Bigl[1,\frac{1}{\sqrt{\Gamma_{\N} D_{N}/v_w^2}} \Bigr]
	\times {\rm min}\Bigl[1, \Gamma_{\LHN} D_{N}/v_w^2 \Bigr]
	\times {\rm exp} \Bigl[ - \int_{0}^{T_{\rm L}} \frac{\ud T}{T} \frac{\Gamma_{\rm w.o}(T)}{H(T)} \Bigr]
	\nn & \quad 
	\times \frac{1}{g_{\ast}} \frac{45}{2\pi^2} \times \frac{2 \gamma_w}{\pi^2} \theta(T) \frac{m_N(T)^3}{T^3} \, {\rm min}\Bigl[ (T\tau)^3 \, , \, 0.1 (T\tau)^{-1} \Bigr] e^{-m_N(T)/T} 
\end{align}
where we have written the entropy density as $s = (2\pi^2/45) g_{\ast} T^3$.  
Note that the dependence on the wall thickness $L_w$ cancels out when we multiply the wall passage time $L_w/v_w$ with the CP-violating phase gradient $d\theta/dz \approx \theta(T) / L_w$.  
In the parameter regime of interest, it is generally the case that $1 \gg \bigl( \Gamma_{\N} D_{N} / v_w^2 \bigr)^{-1/2}$ and $(T\tau)^3 \gg 0.1 (T\tau)^{-1}$.  
For the fiducial parameters we have $1 \sim \Gamma_{\LHN} D_{N}/v_w^2$, but in the regime $1 \gg \Gamma_{\LHN} D_{N}/v_w^2$ a number of factors cancel out, and the expression for the baryon asymmetry simplifies.  
Using the formulas throughout the text we have 
\begin{align}\label{eq:nB_approx}
	\frac{n_{\rm B}}{s} & \approx 
	\bigl( 1 \times 10^{-10} \bigr) 
	\left( \frac{m_N(T_{\rm L})}{10^{11} \GeV} \right) 
	\left( \frac{\theta(T_{\rm L})}{2\pi} \right) 
	\left( \frac{v_w}{0.1} \right)^{-1} 
	\left( \frac{g_{\ast}}{106.75} \right)^{-1} 
	\left( \frac{ x^2 e^{-x} e^{-32.5 \, x^{5/2} e^{-x}} }{4 \times 10^{-3}} \right)
\end{align}
where $x = m_N(T_{\rm L}) / T_{\rm L}$.  
The $x$-dependent factor is maximized at $x \simeq 8.9$ where its value is approximately $4 \times 10^{-3}$.  
Note that the dependence on $\kappa$ has dropped out; this is because the $\kappa$-dependence enters through $D_{N}$ and $\tau$, but $D_{N} \sim \tau^2$.  

\begin{figure}[t]
\begin{center}
\includegraphics[width=0.48\textwidth]{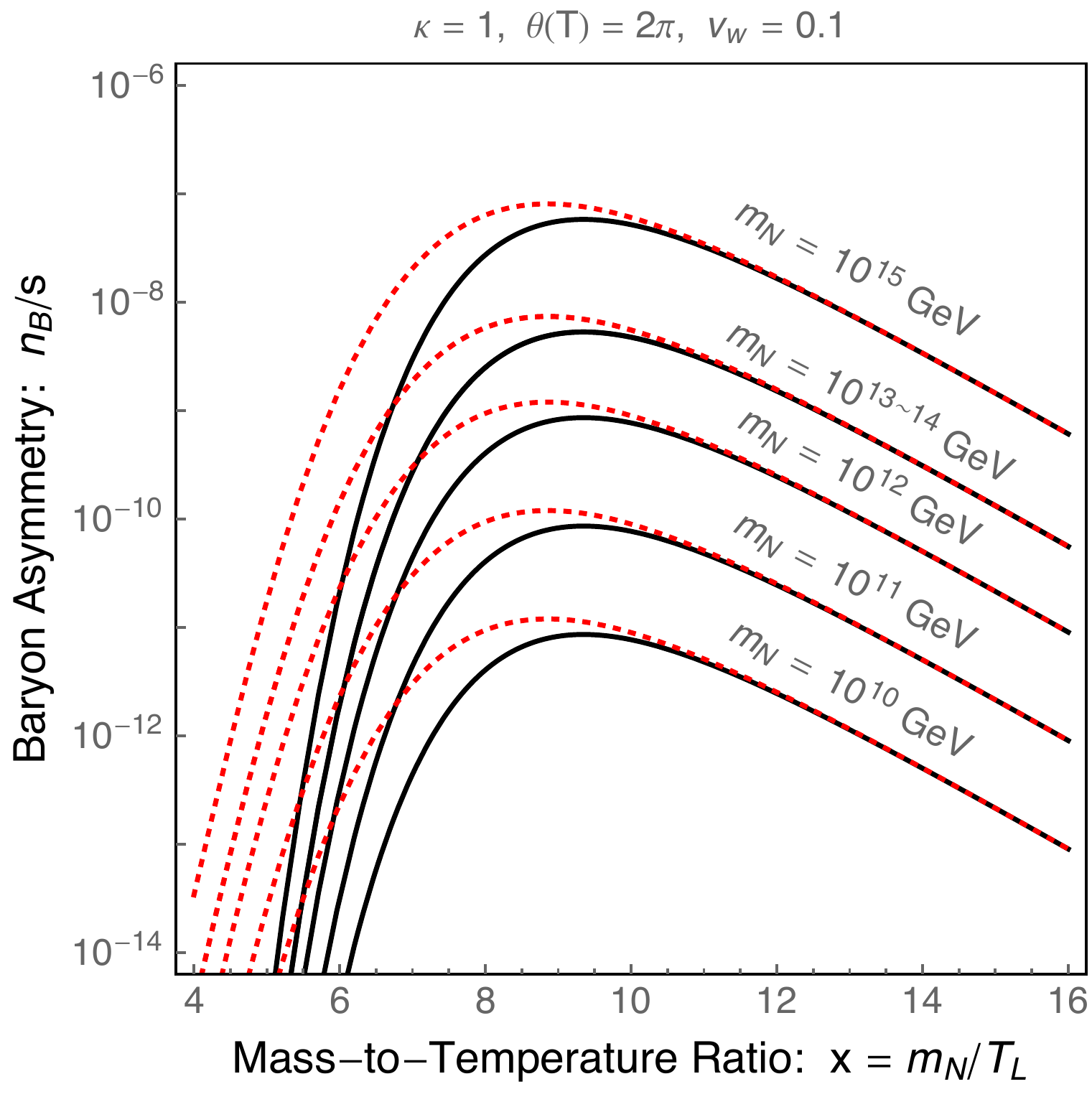} \hfill
\includegraphics[width=0.48\textwidth]{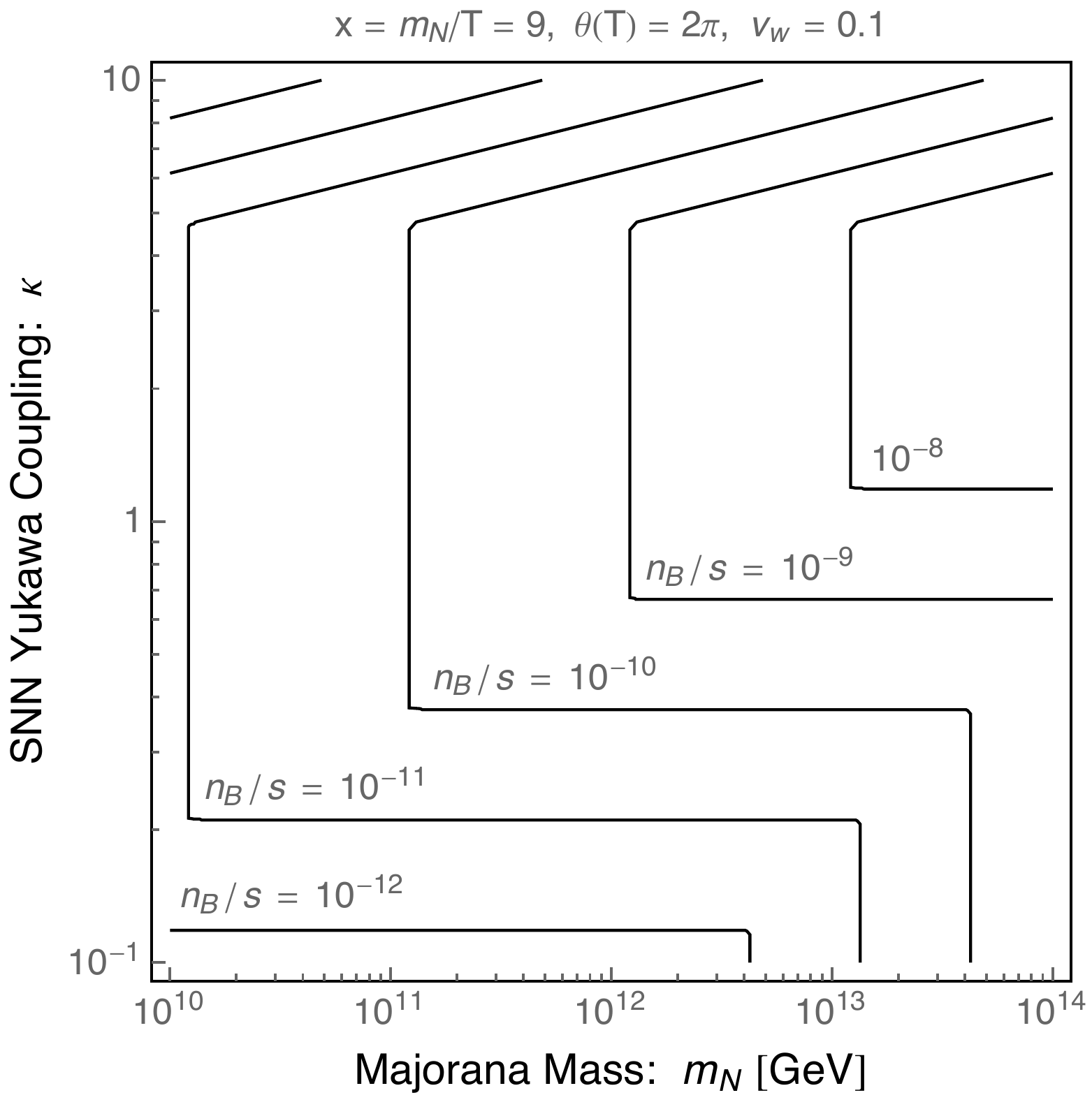} 
\caption{\label{fig:YB_versus_x}
{\it Left:}  The baryon-to-entropy ratio $n_B/s$, expressed as a function of the phase transition temperature $T_{\rm L}$ and the Majorana neutrino mass $m_N$.  
For the black curves we use the exact washout factor from \eref{eq:epsLL}, and for the red (dashed) curves we use the approximation from \eref{eq:epsLL_approx}, which appears in the expression for $n_B/s$ from \eref{eq:nB_approx}. 
{\it Right:}  Variation of $n_B/s$ over the parameter space with $x = m_N / T = 9$ fixed.  
}
\end{center}
\end{figure}

In the left panel of \fref{fig:YB_versus_x} we plot the baryon asymmetry as a function of the phase transition temperature $T_{\rm L}$ and the Majorana neutrino mass $m_N$.  
There is a linear relationship between $n_B/s$ and $m_N$ when $x = m_N / T_{\rm L}$ is held fixed; this can be seen from \eref{eq:nB_approx}.  
In order to accomodate the observed baryon asymmetry of the universe, $n_B / s \simeq 0.9 \times 10^{-10}$ we require the Majorana neutrino mass to be larger than $m_N \approx 10^{11} \GeV$.  
In the right-panel we show the baryon asymmetry as function of $m_N$ and the $SNN$ Yukawa coupling $\kappa$.  
Over much of the parameter space, $n_B/s$ is insensitive to the value of $\kappa$.  
However, for large $\kappa$ we have a power suppression of the source $\sim \tau^3$, that explains the behavior of the isolines at $\kappa \gtrsim 5$.  
In the ``plateau'' region, $n_B/s$ is insensitive to $m_N$ but varies with $\kappa$.  
This regime corresponds to the case where $\sqrt{\Gamma_{\N} D_{N}/v_w^2}>1$ and $1>\Gamma_{\LHN} D_{N}/v_w^2$, so that the two minimum conditions select the combination $\Gamma_{\LHN}\sqrt{D_N}/\sqrt{\Gamma_{\N}}$ that does not depend on $m_N$.

\section{Particle Physics Model}\label{sec:Model}

In this section we discuss a couple of concrete particle physics model that could be used to implement our proposed baryogenesis mechanism.  

\subsection{A weakly coupled model}\label{sec:weaklycoupled}

We have already presented a weakly coupled model in \sref{sub:Overview}.  However, in the model of \eref{eq:L_int}, a single scalar field $S$ is responsible for breaking the $\U{1}_{\rm L}$ symmetry.  
As we discuss here, in order to achieve a CP-violating phase gradient at the bubble wall, we must extend the model to include a second scalar field.  
Furthermore, in order to achieve the correct neutrino mass spectrum, we must extend the model to include at least one additional heavy Majorana neutrino.  

Let the SM be extended to include three left-chiral Weyl spinor fields $N_i$ for $i=1,2,3$ and a pair of complex scalar fields $S_{a}$ for $a=1,2$.  
These fields are charged under $\U{1}_{\rm L}$ as ${\rm L}(N_i) = -1$ and ${\rm L}(S_a) = +2$.  
The SM lagrangian is extended to include 
\begin{align}\label{eq:L_S1S2}
	\Delta \mathscr{L} & = i N_i^{\dagger} \bar{\sigma}^{\mu} \partial_{\mu} N_i + \partial_{\mu} S_a^{\ast} \partial^{\mu} S_a - \Bigl[ \frac{1}{2} \kappa_{ajk} S_a N_j N_k + (\lambda_N)_{ij} L_i H N_j + \hc \Bigr] - U
\end{align}
where a sum over repeated indices is implied, and $U$ is a scalar potential. 
The scalar potential (which does not include the SM Higgs potential) needs to be a polynomial of  the form 
\begin{align}\label{eq:U}
	U(|S_1|^2, |S_2|^2, S_1^* S_2, |H|^2)\supset \mu_1^2 |S_1|^2 + \mu_2^2 |S_2|^2 + \mu_{12}^2 \big[e^{i\delta} S_1^* S_2 + h.c.\big] + \mathrm{quartics}
	\per 
\end{align}
The parameters of $U$ are chosen such that $S_1$ and $S_2$ both acquire vacuum expectation values and the $\U{1}_{\rm L}$ symmetry is spontaneously broken, notice also that the above potential enforce a non-vanishing vacuum expectation value for one CP-odd component of the scalars.  

Let us first discuss the spectrum of light neutrinos.  
The scalar fields acquire vacuum expectation values, $\vev{S_a} = v_{a} / \sqrt{2}$ and $\vev{H} = ( 0 \, , \, v / \sqrt{2})$, and the Yukawa interactions induce masses $\Delta \mathscr{L} = -(1/2) (M_N)_{ij} N_i N_j - (M_D)_{ij} \nu_{L,i} N_j + \hc$ where $(M_N)_{ij} = \kappa_{aij} v_{a} / \sqrt{2}$ is the Majorana mass matrix and $(M_D)_{ij} = (\lambda_N)_{ij} v / \sqrt{2}$ is the Dirac mass matrix.  
Integrating out the heavy neutrinos induces a Majorana mass matrix for the light neutrinos, which is given by the matrix product $M_{\nu} = - M_D M_N^{-1} M_D^{T}$.  
In general the spectrum will contain three massive neutrinos, however in principle only two Majorana right-handed neutrinos are needed to match observations.

Next let us discuss whether the $\U{1}_{\rm L}$-breaking phase transition is first order.\footnote{The phase transition in the closely-related singlet-majoron model has been studied extensively, particularly in association with electroweak symmetry breaking \cite{Kondo:1991jz,Enqvist:1992va,Sei:1992np,Cline:2009sn}.  
Various avenues for achieving a first order phase transition are discussed from a general perspective in \rref{Chung:2012vg}.   }  
The nature of the $\U{1}_{\rm L}$-breaking phase transition depends on the parameters of the scalar potential (\ref{eq:U}) as well as the couplings of the $S_{a}$ to particles in the plasma.  
In general, one should calculate the thermal effective potential $V_{\rm eff}(|S_1|^2, |S_2|^2, S_1^* S_2, |H|^2)$.  
If there is some range of temperature for which the potential exhibits two local minima -- one minimum where $\langle S_1 \rangle = \langle S_2 \rangle = 0$ and a second where both $\langle S_1 \rangle$ and $\langle S_2 \rangle$ are nonzero -- then the phase transition will be first order.  
For example, the barrier may arise from the interactions of $S_a$ with light Higgs bosons in the plasma.  
The effective potential receives a contribution $\Delta V_{\rm eff} = - 4[\tilde{m}_H^2(S_a)]^{3/2} T / 12 \pi$ from the four degenerate components of the Higgs doublet with field-dependent masses $\tilde{m}_H(S_a)$ (lepton number is still conserved).  
In a regime where $\tilde{m}_H^2(S_a) \sim S_a^2$, this contribution to the effective potential is a cubic term, which can induce a barrier in $V_{\rm eff}$ and may drive a first order phase transition.  
We leave this calculation for future work.  

Next we discuss what is needed to obtain a CP-violating phase gradient at the bubble wall.  During the first order $\U{1}_{\rm L}$-breaking phase transition, the scalar field expectation values become inhomogeneous at the bubble wall.  
We can write the field profiles as $\langle S_a \rangle = (v_a(x) / \sqrt{2}) \, e^{i \theta_a}$ for $a=1,2$.  
Energy considerations suggest that $\theta_1$ and $\theta_2$ will be homogenous, since additional field gradients cost energy.  
In the inhomogeneous background of the $S_a$ fields, the Yukawa interactions in \eref{eq:L_S1S2} induce a Majorana mass matrix 
\begin{align}
	M_N(x) & = \left( \kappa_{1} \cos \beta(x) e^{i (\theta_1-\theta_2)} + \kappa_{2} \sin \beta(x) \right) \frac{v_{\rm L}(x)}{\sqrt{2}} 
\end{align}
where we have suppressed the $ij$ flavor indices and identified the physical CP phase $\theta_1-\theta_2$.  
Here we have defined $v_{\rm L}(x) \equiv \sqrt{v_1^2 + v_2^2}$ and $\tan \beta(x) \equiv v_2 / v_1$.  
In this background, the dynamics of $N$ are described by the effective theory, which we discussed previously in \sref{sub:CPV_Phase}.  
The inhomogeneous Majorana mass matrix can be written as $M_N(x) = m_N(x) \, {\rm exp}[i \theta(x)]$ where the physical phase is given by 
\begin{align}\label{theta}
	\theta(x) & = \arctan\bigg[\frac{\kappa_1 \cos \beta(x) \sin (\theta_1-\theta_2)}{\kappa_1 \cos \beta(x) \cos (\theta_1-\theta_2) + \kappa_2 \sin \beta(x) } \bigg]	\per 
\end{align}
It follows that $\partial_{\mu} \theta$ is proportional to $\partial_{\mu} \beta$, which is familiar from studies of electroweak baryogenesis in supersymmetric models (see for example \cite{Carena:1997gx}).  
Therefore, it is clear that two independent contributions to the mass of the right-handed neutrinos are necessary to achieve the phase gradient, which is required for CP-violation.  

In order to calculate the field profiles, $v_a(x)$ and $\theta_a$, a scalar potential $U$ must be specified, and the thermal effective potential must be derived.  
If the phases $\theta_1$ and $\theta_2$ are sampled uniformly from the interval $[0,2\pi)$ then the phase gradient will be positive for some bubbles and negative for others.  
Consequently, the global lepton-number will remain equal to zero even though individual bubbles develop an excess of either leptons or anti-leptons.  
To avoid this outcome, it is necessary that the scalar potential $U$ contains CP-violating phases that bias $\theta_1 - \theta_2$ to a preferred, nonzero value.  

\subsection{A strongly coupled model}\label{sec:stronglycoupled}

Confining gauge theories present an elegant framework for achieving a first order phase transition.  
In the presence of fundamental fermions the confining phase transition can spontaneously break the chiral symmetries associated to the light flavors.
In particular in $\SU{3}$ gauge theories (see for example \cite{Svetitsky:1982gs,Pisarski:1983ms,Aoki:2006br}), the confining phase transition is first order when at least three flavors, $\psi_i$ and $\psi^c_i$ with $i=1,2,3$, are sufficiently lighter than the confinement scale. Here $\psi$ and $\psi^c$ are a 3 and a $\bar{3}$ of the new $\SU{3}$ confining group, and they are singlets under the SM.

In order to match onto our model of baryogenesis, one of the chiral symmetries should correspond to lepton-number, $\U{1}_{\rm L}$.  
One can envision a model such as 
\be
\Delta \mathscr{L}\supset \kappa_{ij} \frac{\psi_i \psi^c_j}{\Lambda_{\rm UV}^2} N N + \lambda S NN + c_{ij} S^* \psi_i \psi_j^c + \hc
\ee
where we have written all the renormalizable interactions for $N$ and $\psi$ allowed by the gauge symmetry and $\U{1}_{\rm L}$, and we have written the leading higher-dimensional operator, which is needed to generate the right-handed neutrino mass. 
The right-handed neutrino $N$ gets mass from the ``techni-color'' condensate $\langle \psi_i \psi_j^c \rangle$ as well as a weakly coupled source $\langle S \rangle$, which is needed to get a non-trivial CP-violating gradient.  
When the condensate forms, $\langle \psi_i \psi^c_j \rangle \sim \delta_{ij} \Lambda^3/(16\pi^2)$, it will also induce a tadpole for $S$ at approximately the same scale. As shown in the previous section one needs a misalignment between the two ($z$-dependent) sources of lepton-number breaking and a physical CP-violating phase arises as long as the ratio $\Lambda(x)^3/\langle S \rangle$ depends on the spacetime coordinate $x$.

Since the scale of lepton-number violation is typically $m_N \sim 10^{12}\GeV$ in our model, it would be interesting to explore the possible relation with (composite) axion models.  
Similarly to those models, here massless fundamental fermions are required, since fermion masses, $m \psi \psi^c$, can explicitly break lepton-number.

\section{Phenomenology Highlights}\label{sec:Pheno}

Here we discuss a few aspects of the phenomenology.  

\subsection{Neutrinoless Double Beta Decay}\label{sub:0nbb}

Since the neutrinos are Majorana particles, the neutrinoless double beta decay ($0\nu \beta \beta$) channel is not forbidden by any conservation law.  
For a recent review see \rref{DellOro:2016tmg}.  
The $0\nu \beta \beta$ rate is proportional to the squared effective mass $m_{\beta \beta} \equiv \bigl| \sum_i U_{ei}^2 m_i \bigr|$.  
The next generation of $0\nu \beta \beta$ experiments expects to reach a sensitivity of $\sigma(m_{\beta \beta}) \sim (100-200) \meV$.  

\subsection{Majoron}\label{sub:Majoron}

Since the global $\U{1}_{\rm L}$ symmetry is spontaneously broken, the spectrum contains a massless Goldstone boson; this is the so-called majoron \cite{Chikashige:1980ui, Gelmini:1980re}.  
At low energies the heavy leptons, $N$ and $S$, have been integrated out of the theory, and we are interested in the interactions between the pseudoscalar majoron field $J$ and the SM leptons $L_i$ and $E_i$.  
To make these interactions evident, let us perform the field redefinition, $L_i \to L_i \, e^{iJ /2 v_{\rm L}}$ and $E_i \to E_i \, e^{-iJ /2 v_{\rm L}}$, which follows from the $\U{1}_{\rm L}$ charge assignments.  
Thus, the majoron acquires a derivative interaction 
\be\label{eq:J_int}
	\mathscr{L}_{\rm int} = - \frac{\partial_\mu J}{2 v_{\rm L}} j_{\rm L}^\mu 
\ee
where $j_{\rm L}^{\mu} = L_i^{\dagger} \bar{\sigma}^{\mu} L_i - E_i^{\dagger} \bar{\sigma}^{\mu} E_i$ is the SM lepton-number current.  
The interaction is put into a more convenient form if we first integrate by parts to obtain $\mathscr{L}_{\rm int} = (J/2v_{\rm L}) \partial_{\mu} j^{\mu}_{\rm L}$.  
The lepton-number current is not conserved, but rather it is violated both explicitly by the Majorana mass and anomalously by the SM weak interactions:  $\partial_{\mu} j_{\rm L}^{\mu} = ( \lambda_N^2 LHLH / m_N + \hc) + 3 (\alpha_w/8\pi) W \tilde{W} - 3(\alpha_y/8\pi) B \tilde{B}$.   
After electroweak symmetry breaking, $\vev{H} = ( 0 \, , \, v / \sqrt{2})$, the interaction of the majoron with the SM neutrinos becomes 
\begin{align}\label{eq:L_Jnunu}
	\mathscr{L}_{\rm int} \supset - \frac{i}{2} g_{J\nu\nu} J \nu \nu + \hc 
\end{align}
where $g_{J \nu \nu} \equiv - m_{\nu} / v_{\rm L}$.  

The majoron-neutrino Yukawa interaction (\ref{eq:L_Jnunu}) leads to an array of well-studied phenomenology.  
Majorons may be produced in stellar environments, and limits on supernoave cooling impose an upper bound on the Yukawa coupling $g_{J \nu\nu}$ at the level of $10^{-7}$ to $10^{-5}$ for different flavor components \cite{Kachelriess:2000qc, Farzan:2002wx}.  
Comparable bounds arise from anomalous meson and lepton decays into majorons \cite{Lessa:2007up}.  
However, for the parameters of interest $g_{J \nu \nu} \sim 10^{-22} ( v_{\rm L} / 10^{12} \GeV)^{-1}$, and these bounds are easily evaded.  

The couplings to electrons (and quarks) arises at the one-loop through interactions with the $W$ and $Z$ bosons. The contribution (neglecting off-diagonal flavor mixing) is proportional to \cite{Chikashige:1980ui}
\be
	g_{Jee}\simeq \frac{\lambda_N^2}{8\pi^2}\frac{m_e}{v_L}\sim \frac{\kappa}{8\pi^2} \frac{m_\nu m_e}{v^2} \sim \kappa\, 10^{-20},
\ee
while the coupling to quarks is obtained replacing $m_e \to m_q$.  
The majoron's coupling with SM matter is also sensitive to the explicit breaking of the lepton-number.  
For example, a mixing between the majoron and the SM Higgs gives rise to new interactions.  

\subsubsection{Majoron Mass}\label{sub:Mass}

In the model considered here, the $\U{1}_{\rm L}$ symmetry is not broken explicitly, and the majoron Goldstone boson is exactly massless.  
However, higher-dimensional operators may break $\U{1}_{\rm L}$ and contribute to the mass of the majoron.  
For instance, the operator $c_{\slashed{L}}S |H|^4/ M_{\rm Pl}$ induces a majoron mass $m_J \sim v^2 /\sqrt{\Mpl v_{\rm L}} \sim \bigl( 0.01 \eV \bigr) \sqrt{c_{\slashed{L}}(10^{12} \GeV/v_{\rm L})}$.  

If $\U{1}_{\rm L}$ is broken explicitly, this may threaten to disrupt the baryogenesis mechanism.  
Specifically, the operator $S |H|^4$ opens new channels for lepton-number-violating washout, such as $S H H \leftrightarrow HH$.  
We estimate the rate for these $\Delta \mathrm{L}=2$ processes as $\Gamma_{\slashed{L}}(T)\sim c_{\slashed{L}}^2 T^3/M_{\rm Pl}^2$.  
Provided that $c_{\slashed{L}} \sim O(1)$, the new washout process is out of equilibrium, $\Gamma_{\slashed{L}}(T) \ll H(T)$, and it can be safely neglected.  

\subsubsection{Majoron as Dark Radiation}\label{sub:Dark_Rad}

The interactions in \eref{eq:J_int} keep the majoron in thermal equilibrium at high temperature.  
We estimate the interaction rate as $\Gamma \sim \lambda_N^4 T^3 / m_N^2$, which can be written as $m_{\nu}^2 T^3 / v^4$ using \eref{eq:lamN}.  
The interaction rate drops below the Hubble expansion rate, $H \sim T^2 / (10 \Mpl)$, at temperatures $T \lesssim 10^{10} \GeV$.  
At this time, the majoron particles decouple from the thermal bath.  

In our model the majoron is very light (possibly massless) and very long-lived.  
Consequently, the relic abundance of relativistic majoron particles will contribute to the effective radiation density of the universe.  
Since the majorons decouple so early in the cosmological history, they do not receive the entropy injections from the decoupling of the other SM species.  
As a result, the relic majoron background is colder than the relic neutrino background by a factor of $\sim 1 / g_{\ast}^{1/3} \simeq 0.2$, and the corresponding contribution to the effective number of neutrino species is $\Delta N_{\rm eff} \approx 0.027$ \cite{Baumann:2016wac}.  
The relic majoron background evades current CMB limits on additional radiation density, but the improved sensitivity of the CMB Stage-IV telescopes may be able to pick up this subtle effect \cite{Abazajian:2016yjj}.  

\subsubsection{Majoron as Dark Matter}\label{sub:DM}

If the majoron is massive, as we discussed above, then it provides a dark matter candidate.  
Depending on neutrino mass spectrum, the majoron may be unstable toward the decay into a pair of light neutrinos via the interaction in \eref{eq:L_Jnunu}.  
For the fiducial scales considered above, $v_{\rm L} \sim 10^{12} \GeV$ and $m_J \sim 0.01 \eV$, the Majoron lifetime greatly exceeds the age of the universe today, and it is effectively stable.  

Although majorons decouple early from the thermal bath, they can be produced non-thermally from the misalignment mechanism \cite{Preskill:1982cy, Abbott:1982af, Dine:1982ah,Rothstein:1992rh}.  
The Hubble friction will become subdominant at a temperature $T_J \sim \sqrt{m_J \Mpl} \bigl( \pi^2 g_{\ast}(T_J) / 10 \bigr)^{-1/4}$ when $m_J\sim 3 H(T_J)$, and after that the Majoron will start to oscillate around the minimum of its potential, much similar to what happens to axion DM  (for a review on the cosmology of light pseudoscalars see \cite{Kim:1986ax}).   \footnote{Oscillations of majoron field can also be used to generate a lepton asymmetry \cite{Ibe:2015nfa}.  } The energy density of the oscillations will behave as cold dark matter.  
The yield today is estimated as 
\be
	\frac{n_J}{s} \sim \langle \theta_J^2 \rangle \, \frac{m_J v_{\rm L}^2}{s(T_J)} 
\ee
where we average over different initial misalignment angles.  
The present energy density is estimated as
\begin{align}
	\Omega_J 
	= \frac{m_J \, n_J(T_0)}{3 \Mpl^2 H_0^2} 
	\sim \frac{\langle \theta_J^2 \rangle m_J^2 v_{\rm L}^2}{3 \Mpl^2 H_0^2} \frac{g_{\ast S}(T_0) T_0^3}{g_{\ast S}(T_J) T_J^3}
	\sim \left( \frac{\pi^{3/2}}{3 \times 10^{3/4}} \frac{g_{\ast S}(T_0) T_0^3}{\Mpl^{7/2} H_0^2} \right) \frac{\langle \theta_J^2 \rangle m_J^{1/2} v_{\rm L}^2}{g_{\ast}(T_J)^{1/4}} 
\end{align}
where $H_0 \simeq 2 \times 10^{-42} \GeV$ is the Hubble constant, $T_0 \simeq 2.34 \times 10^{-13} \GeV$ is the temperature of the CMB, $g_{\ast S}(T_0) \simeq 3.91$, and $g_{\ast S}(T_J) \approx g_{\ast}(T_J) \simeq 106.75$ is the effective number of relativistic species.  
In the last line, we have used the expression for $T_J$ from above.  
As we discussed in \sref{sub:Mass}, the majoron mass depends on the operator responsible for explicit lepton-number violation, and we therefore treat $m_J$ as a free parameter.  
Therefore the majoron DM relic abundance is given by 
\be
	\Omega_J  \sim 0.2 \, \Big( \frac{\langle \theta_J^2\rangle}{\pi^2} \Big) \, \Bigl( \frac{m_J}{0.1 \meV} \Bigr)^{1/2}  \, \Bigl( \frac{v_{\rm L}}{10^{12} \GeV} \Bigr)^{2} 
	\per
\ee
This agrees well with the observed relic abundance of dark matter, $\Omega_{\rm DM} \sim 0.2$. 
Notice also that the dynamics of the strings and domain walls can affect the contribution to the energy density of Majorons (see subsection \ref{sub:Cos_String}).

\subsection{Gravitational Wave Background}\label{sub:Grav_Wave}

An essential ingredient in our baryogenesis mechanism is that the $\U{1}_{\rm L}$-breaking phase transition is first order.  
A first order cosmological phase transition also leads to the production of gravitational waves \cite{Hogan:1986qda, Kamionkowski:1993fg}.  
Therefore, the existence of a stochastic gravitational wave background is an inevitable secondary prediction of our mechanism.  

Gravitational radiation arises partially from the collision of bubbles and partially from the decay of turbulence and sound waves in the plasma.  
Since there are multiple source of gravitational waves, predictions for the spectrum of gravitational radiation are very model-dependent.  
However, it is a general prediction that the spectrum is peaked at an intermediate frequency $f_p$, and the value of this frequency can be inferred robustly in terms of the phase transition temperature, because it is related to the size of the cosmological horizon at the time of the phase transition.  
Assuming that the bubbles collide when their diameter is a fraction $x$ of the cosmological horizon, we have $f_p \approx (10^{5} \Hz) x^{-1} (T_{\rm L} / 10^{11} \GeV)$.  

The stochastic background of gravitational wave radiation will be probed by gravitational wave interferometers such as LIGO \cite{TheLIGOScientific:2016dpb} and LISA \cite{Caprini:2015zlo}.  
The sensitivity of LIGO peaks at $f \sim 10^2 \Hz$ and the sensitivity of LISA peaks at $f \sim 10^{-3} \Hz$.  
Therefore LIGO or a future high-sensitivity interferometer like BBO \cite{Harry:2006fi} or DECIGO \cite{Kawamura:2011zz} may be best equipped to search for the high-frequency gravitational wave radiation produced during the first order $\U{1}_{\rm L}$-breaking phase transition.  

\subsection{Cosmic String Network}\label{sub:Cos_String}

It is a necessary ingredient in our model that the $\U{1}_{\rm L}$ symmetry is spontaneously broken through a cosmological phase transition.  
In general, cosmic strings will form during a phase transition in which a $\U{1}$ symmetry becomes broken \cite{Kibble:1976sj, Zurek:1985qw}.  
The subsequent evolution of the cosmic string network depends on whether the $\U{1}$ symmetry was global or gauged, and whether it was also explicitly broken.  

If the $\U{1}_{\rm L}$ symmetry is global and not explicitly broken in the lagrangian, then the network of topological defects is made up of global strings \cite{Vilenkin:1982ks}.  
When string loops are pinched off from the network of long strings, they efficiently radiate Goldstone bosons (massless majorons) thereby damping the high frequency oscillation modes of the string loop \cite{Battye:1993jv} and suppressing the gravitational wave radiation \cite{Battye:1997ji}.  
However, the majoron emission may provide an additional non-thermal component of dark radiation, which is discussed in \sref{sub:Dark_Rad}.  
A scale-invariant spectrum of stochastic gravitational waves arises as long strings enter the horizon and the scalar field has to self-order \cite{Figueroa:2012kw}.  
The gravitational wave radiation may be within reach of future space-based gravitational wave interferometers, such as BBO \cite{Harry:2006fi} and DECIGO \cite{Kawamura:2011zz}.  

If the $\U{1}_{\rm L}$ symmetry is global and broken explicitly, then the topological defect network consists of strings connected by domain walls \cite{Vilenkin:1982ks}.  
If the $\U{1}_{\rm L}$-breaking term is a linear, such as $S |H|^2$ or $S |H|^4$, then the topological defects can decay.  
Specifically, strings become connected by domain walls, and the tension of the wall causes the configuration to collapse while losing energy into the radiation of pseudo-Goldstone bosons (massive majorons) \cite{Chang:1998tb}.  
On the other hand, if the $\U{1}_{\rm L}$ symmetry is broken by an operator which leaves a $\Zbb_n$ discrete subgroup, such as $S^2$ or $S^3$, then the domain walls are stable.  
This is not a cosmologically viable scenario, as the domain wall energy density will eventually come to dominate \cite{Hiramatsu:2012sc}.  

Finally, if the $\U{1}_{\rm L}$ symmetry is gauged, then the defect network is composed of gauge (or Abelian-Higgs) strings.  
In fact, since the $\U{1}_{\rm L}$ symmetry is anomalous, we should consider instead gauged $\U{1}_{\rm B-L}$ strings, which can arise in models of grand unification \cite{Jeannerot:1995yn}.  
For the high symmetry breaking scales that we consider here, the primary energy loss mechanism\footnote{For lower values of the string tension, the emission of SM Higgs bosons can also be significant \cite{Long:2014lxa}.} is the radiation of gravitational waves, which is not very efficient, and therefore the string loops are long-lived.  
The presence of a cosmic string network in the universe today generates a stochastic background of gravitational wave radiation as string loops oscillate and periodically form cusps where gravitational wave radiation is enhanced.  
The low frequency gravitational wave background is constrained by observations of pulsar timing.  
These limits can be expressed as $G\mu / c^2 \lesssim 2.8 \times 10^{-9}$ \cite{Blanco-Pillado:2013qja} where $G \simeq (1.22 \times 10^{19} \GeV)^{-2}$ is Newton's constant and $\mu$ is the string tension.  
Typically $\mu$ is set by the scale of symmetry breaking, which we have denoted as $v_{\rm L}$ for the $\U{1}_{\rm L}$-breaking phase transition.  
In terms of the symmetry breaking scale, $\mu \sim v_{\rm L}^2$, the pulsar timing limit becomes $v_{\rm L} \lesssim 6.5 \times 10^{14} \GeV$.  
Therefore, models with large $m_N = \kappa v_{\rm L} / \sqrt{2}$, which are favorable for baryogenesis, may be constrained by the non-observation of stochastic gravitational waves arising from the cosmic string network.  
While we do not expect that gauging the $\U{1}_{\rm L}$ symmetry will dramatically affect the dynamics of baryogenesis, this scenario surely merits further investigation.

\subsection{Reheating after Inflation}\label{sub:Reheating}

As we have seen in \sref{sub:BaryonAsym}, in order for the predicted baryon asymmetry to match the observed value, we need the Majorana mass scale to be large, for instance $m_N \gtrsim 10^{12} \GeV$.  
Since the baryon asymmetry is maximized for $x = m_N / T_{\rm L} \sim 10$ this implies a lower bound on the temperature of the $\U{1}_{\rm L}$-breaking phase transition, namely $T_{\rm L} \gtrsim 10^{11} \GeV$.  
Since the plasma forms at the end of inflation during reheating, we therefore impose a lower bound on the reheat temperature $T_{\rm RH} \gtrsim 10^{11} \GeV$.  
Our model of baryogenesis is most naturally accommodated in models of high scale inflation or models with efficient reheating.  

\section{Discussion}\label{sec:Discussion}

In this article, we have proposed a new model of baryogenesis from leptogenesis, which relies upon a first order $\U{1}_{\rm L}$-breaking phase transition.  
The lepton asymmetry is generated by the CP-violating scattering of neutrinos at the bubble wall, and in this regard the model shares many common features with electroweak baryogenesis.  
We have estimated the resultant baryon asymmetry in \sref{sec:Baryogen}.  
A more accurate prediction could be made with improvements to the source and transport calculations, but we do not expect that the qualitative results will be changed.  

Although lepton-number is violated inside of the bubbles, washout is avoided provided that the phase transition is strongly first order.  
That is to say, the ratio of the Majorana neutrino mass and the phase transition temperature, $m_N / T_{\rm L}$, should be sufficiently large to suppress lepton-number-violating scattering among the SM leptons and Higgs bosons.  
At the same time, the mass cannot be too large, otherwise it becomes energetically disfavored for the Majorana neutrinos to enter the bubbles.
These two conditions bracket the phase transition temperature $T_{\rm L}$ to satisfy $m_{N} / T_{\rm L} \sim 10$, and the baryon asymmetry is exponentially suppressed for either larger or smaller temperatures.  
In this work we have taken $T_{\rm L}$ as a free parameter, but it would be interesting to perform a full phase transition study on the model in \sref{sec:Model}.  

The amplitude of the baryon asymmetry is suppressed by $\lambda_N^2$, that is the squared Yukawa coupling associated with the $LHN$ interaction.  
This coupling controls the efficiency with which $N$-number is converted into $L$-number in front of the bubble wall.  
The seesaw relation (\ref{eq:lamN}) relates $\lambda_N$ to the Majorana mass scale $m_N$, and consequently the baryon asymmetry is suppressed as we lower the scale of lepton-number violation, as we see in \fref{fig:YB_versus_x}.  
It may be interesting to explore other seesaw scenarios or a nontrivial flavor structure in order to break the naive seesaw relation and thereby achieve the desired baryon asymmetry even for a lower Majorana mass scale.  

The model presented here draws upon some of the most appealing features of leptogenesis and electroweak baryogenesis.  
The model inherits its connection with neutrino physics from leptogenesis, which restricts the number of free parameters by predicting relations with the spectrum of light neutrinos.  
Similar to electroweak baryogenesis, the model admits a number of interesting cosmological probes associated with the first order phase transition.  
These include a stochastic background of gravitational wave radiation, a network of cosmic strings, relativistic bath of dark radiation, and a dark matter candidate.  
The detection of these ``baryogenesis by-products'' will be challenging, but if the endeavor is successful, then future cosmological observations may point the way toward understanding the origin of the matter / antimatter asymmetry.  

\subsubsection*{Acknowledgements}
We are grateful to Carlos Wagner for illuminating discussions, and we thank Bjorn Garbrecht for valuable comments on the manuscript.  
AJL is supported at the University of Chicago by the Kavli Institute for Cosmological Physics through grant NSF PHY-1125897 and an endowment from the Kavli Foundation and its founder Fred Kavli.  
AT acknowledges support by an Oehme Fellowship.  
LTW is supported by DOE grant DE-SC0013642. 

\begin{appendix}

\section{Derivation of the diffusion transport equations}\label{app:transport}

We are interested in computing the number density of baryons (and leptons).  
This requires solving Boltzmann and diffusion equations for the phase space distribution function of a species $a$, which we denote by $f_a({\bm p},T)$.  

Let $f_a({\bm x}, {\bm p},T)$ be the phase space distribution function of species $a$, and let $n_a({\bm x}, T)$ be the number density of species $a$.  
The number density is evaluated as 
\be\label{eq:na_def}
	n_{a} 
	= g_{a} \int \! \! \frac{\ud^3 {\bm p}}{(2\pi)^3} \, f_{a}
\ee
where $g_a$ counts the degrees of freedom.  
If species $a$ is kept in kinetic equilibrium, then $f_{a}$ is well-approximated by the Bose-Einstein or Fermi-Dirac distribution function:  
\be
	f_a 
	= \frac{1}{e^{(E_{a}-\mu_a)/T} \pm 1} 
\ee
where $E_a = \sqrt{|{\bm p}|^2 + m_a^2}$ is the energy, $\mu_a(T)$ is the chemical potential, and the $+$ ($-$) is for fermions (bosons).  

For a non-relativistic species, $E_{a} / T > m_a / T \gg 1$, we can approximate 
\be\label{eq:fa_approx}
	f_a \approx \bar{f}_a \, e^{\mu_a/T}
	\qquad \text{and} \qquad 
	n_{a} \approx \bar{n}_a \, e^{\mu_a/T} 
\ee
where $\bar{f}_a({\bm p}, T) = e^{-E_a/T}$ and 
\begin{align}\label{eq:nbar_nonrel}
	\bar{n}_a(m_a \gg T) = g_a \frac{m_a^2 T}{2\pi^2} K_2\bigl(m_a/T\bigr)
	\per 
\end{align}
For a relativistic species, $m_a / T \ll 1$, with a small departure from kinetic equilibrium, $\mu_a/T \ll 1$, we can approximate 
\begin{align}\label{eq:nbar_rel}
	n_a(m_a\ll T)^{\rm bosons} = g_a \frac{\zeta(3)}{\pi^2} T^3 
	\quad , \quad 
	n_a(m_a \ll T)^{\rm fermions} = g_a \frac{3}{4}\frac{\zeta(3)}{\pi^2}T^3.
\end{align}
where $\zeta(x)$ is the zeta function.  
In the following, we will write $f_a$ as in \eref{eq:fa_approx}.  
For non-relativistic particles ($N$ and $\bar{N}$) this is a good approximation. 
For relativistic particles ($L_i$, $\bar{L}_i$, $H$, and $\bar{H}$) we expect the error to be no more than an $O(1)$ factor, since the scattering amplitudes are free from IR divergences.  

The evolution of $n_a$ with time is described by a Boltzmann equation.  
Suppose that particles of species $a$ participate in interactions that change the number of particle of species $a$ by $\delta_a$ units.  
The Boltzmann equations can be written as
\be\label{eq:boltzmann}
\dot{n}_a + 3 H n_a - D_a \nabla^2 n_a = -\sum_{\rm all\ processes} \delta_a \gamma_{a i j\cdots \leftrightarrow k l \cdots} \left( \frac{n_a}{\bar{n}_a}\frac{n_i}{\bar{n}_i}\frac{n_j}{\bar{n}_j} \cdots - \frac{n_k}{\bar{n}_k}\frac{n_l}{\bar{n}_l}\cdots \right) + S_a
\ee
where $H$ is the Hubble parameter, $D_a$ is the diffusion coefficient of species $a$, and $S_a$ is a density-independent source.  
The transport coefficients $\gamma_{a i j\cdots \leftrightarrow k l \cdots}$ are defined as
\be\label{eq:gamma_def}
	\gamma_{a i j\cdots \leftrightarrow k l \cdots} = \sum \! \! \int \ud\Pi_a \bar{f}_a \ud\Pi_i \bar{f}_i \ud\Pi_j \bar{f}_j \cdots \int \ud\Pi_k \ud\Pi_l \cdots \big|\mathcal{A}_{a i j\cdots \to k l \cdots}\big|^2 (2\pi)^4 \delta^{(4)}(\sum p),
\ee
where we sum the initial and final spin states and properly account for identical particles.  
The Lorentz-invariant phase space volume elements are defined as $\ud\Pi_i = \ud^3 {\bm p}_i / (2\pi)^3 / (2 E_i)$.  
In writing \eref{eq:boltzmann} we have assumed that the interactions respect time-reversal invariance, or equivalently both CP- and CPT-invariance.  

We make the following simplifications.  
In calculating the diffusion of charges away from the bubble wall, the time scales of interest are much shorter than $H^{-1}$, and therefore we can drop the Hubble drag term $3Hn_a$ from \eref{eq:boltzmann}.  
Similarly the change in the plasma temperature is negligible on the time scales of interest, {\it i.e.} $\dot{T} = - HT$, and we can treat $T$ as a constant.  
Since particles can acquire mass at the bubble wall, the equilibrium distribution $\bar{n}_a$ may depend on the spatial coordinate.  
We assume that the change in $\bar{n}_a$ is smooth from outside to inside the bubble and that we can neglect derivatives on $\bar{n}_a$. 
Finally we assume that the departures from chemical equilibrium are small, $\mu_a / T \ll 1$, which is an excellent approximation for baryogenesis since the observed baryon asymmetry of the universe corresponds to $\mu/T \sim 10^{-8}$.  
With these assumptions, \eref{eq:boltzmann} simplifies to 
\be\label{eq:boltzmann_2}
	\dot{\mu}_a - D_a \nabla^2 \mu_a = -\sum_{\rm all\ processes} \delta_a \frac{\gamma_{a i j\cdots \leftrightarrow k l \cdots}}{\bar{n}_a}\left( \mu_a + \mu_i + \mu_j +\cdots - \mu_k - \mu_l - \cdots \right) + \frac{S_a}{\bar{n}_a} 
	\per
\ee
In general $\mu_a$ is a function of space ${\bm x}$ and time $t$.  

Now we discuss the various interactions that are relevant to our model of baryogenesis.  
\vspace{-0.2cm}
\begin{itemize}
\item $LHN$ Yukawa-mediated interactions. They play an important role both outside and inside the bubble. Outside they redistribute the excess in $N$ by converting it into an excess of $L$. Inside the bubble they need to be strongly out of equilibrium, so that processes like $LH\to N$ do not erase the excess of $L$-number when it diffuses into the bubble. We define the following transport coefficients (thermally averaged rates per unit volume) 
\begin{eqnarray}
\gamma_0&=&\gamma_{H \leftrightarrow \bar{L}\bar{N}}+\gamma_{L \leftrightarrow \bar{H}\bar{N}}+\gamma_{N \leftrightarrow \bar{L}\bar{H}}= \gamma_{\bar H \leftrightarrow L N}+\gamma_{\bar L \leftrightarrow HN}+\gamma_{\bar N \leftrightarrow LH}\,,\\
\frac{\gamma}{4}&=&\gamma_{N\leftrightarrow \bar{L}\bar{H}}=\gamma_{\bar N\leftrightarrow LH}=\gamma_{N\leftrightarrow LH}=\gamma_{\bar N\leftrightarrow \bar{L}\bar{H}} 
	\com
\end{eqnarray}
which depend on ${\bm x}$ and $t$ in general.  The lepton $L^i$ carries a flavor index ($i=1,2,3$), which is also suppressed when we write $\gamma_0$ and $\gamma$.  The first transport coefficient, $\gamma_0$, describes the re-equilibration of lepton number outside the bubble. In principle all three channels can contribute to $\gamma_0$, but if the Higgs has the largest thermal mass, $m_H > m_L + m_N$, then only the Higgs decay channel will be kinematically open. The second transport coefficient, $\gamma$, describes transitions inside the bubble where $N$ is Majorana.   We neglect any CP-violating effects.  This is different from thermal leptogenesis where CP-violation in the decay of right-handed neutrinos plays a major role.
\item $N$-number violating interactions. Inside the bubble, the right-handed neutrinos pick up a Majorana mass, which tends to erase the asymmetry between $N$ and $\bar{N}$. The rate is indicated as $\gamma_{\Delta N}$ and is mainly due to $N$-number scatterings with the condensates. In the main text we estimated $\Gamma_{\N}= \gamma_{\Delta N}/\bar{n}_N\sim m_N^2/(10 T)$.
\item $\mathrm{L}$-number violating interactions.  In presence of massive right-handed neutrinos, $\Delta \mathrm{L}=2$ transitions are mediated by the Weinberg operator $LHLH$. However the off-shell contribution correspond to interactions at $O(\lambda_N^2)$ that we neglect. The $\mathrm{L}$-breaking interactions are already taken into account by $N$-decays and inverse decays (described by the rate $\gamma$). 
\item Other Yukawa interactions. The $N$-number asymmetry is eventually redistributed to the other SM species via the SM Yukawa interactions.  We indicate the corresponding transport coefficients by $\gamma_{\mathrm{E}^{ij}}$, $\gamma_{\mathrm{U}^{ij}}$ and $\gamma_{\mathrm{D}^{ij}}$ for the charged lepton and quark interactions.  We let $\gamma_{\mathrm{S}}$ indicate the yukawa interaction rate between $N$ and $S$.
\item  Sphaleron transitions.  The weak sphaleron process is slow compared to the diffusion time scale.  Thus we can drop it from the diffusion equations, and account for its effect after the phase transition has completed. Doing so, we neglect a possible contribution to left-handed quark asymmetries during the phase transition, but this will not affect our final results significantly.  We also neglect the strong sphaleron process, which is expected to have an $O(1)$ effect on the quark asymmetries.  
\end{itemize}

Now we write down the Boltzmann equations that are relevant for baryogenesis.  
For each particle species $a$ we denote the corresponding CP-conjugate anti-particle species by $\bar{a}$, and we define $n_{\aaa} = n_a - n_{\bar{a}}$.  
Using \eref{eq:fa_approx} we have the approximation $n_{\aaa}=\bar{n}_a (\mu_a -\mu_{\bar{a}})/T$, which is reliable for $\mu/T \ll 1$.  
Including each of the processes described above, we construct the Boltzmann equations from \eqref{eq:boltzmann_2}
\begin{eqnarray}\label{eq:boltzmann-model}
\dot{n}_{\NNN} +3H n_{\NNN}- D_N \nabla^2 n_{\NNN} &=& -  \sum_i\gamma_0^i \, \Big( \frac{n_{\NNN}}{\bar n_N} + \frac{n_{\LLL^i} }{\bar{n}_{L^i}} + \frac{n_{\HHH}}{\bar{n}_{H}}\Big) -  \gamma_S  \Big( 2 \frac{n_{\NNN}}{\bar n_N} + \frac{{n}_{\Delta S}}{ \bar n_{S}}\Big)\,\nonumber\\
&-& \Big(\sum_i\gamma_i/2+\gamma_{\Delta N}\big)\frac{n_{\NNN}}{\bar{n}_N} + S_{\NNN}\,\nonumber\\
\dot{n}_{\LLL^i}  +3H n_{\LLL^i}- D_L \nabla^2 n_{\LLL^i} &=& - \gamma_0^i \, \Big( \frac{n_{\NNN}}{\bar n_N} + \frac{n_{\LLL^i} }{\bar{n}_{L^i}} + \frac{n_{\HHH}}{\bar{n}_{H}}\Big)  - \frac{\gamma_i}{2}\Big(\frac{n_{\LLL^i}}{\bar{n}_{L^i}} + \frac{{n}_{\HHH}}{ \bar n_{H}}\Big)\nonumber\\ 
&-& \sum_{j}\gamma_{\mathrm{E}^{ij}}\,  \Big( \frac{n_{\EEE^j}}{\bar{n}_{E^j}} + \frac{n_{\LLL^i}}{\bar{n}_{L^i} }- \frac{n_{\HHH}}{\bar{n}_H}\Big)\,,\nonumber\\
\dot{n}_{\HHH}  +3H n_{\HHH}- D_H \nabla^2 n_{\HHH} &=& - \sum_i \gamma_0^i \, \Big( \frac{n_{\NNN}}{\bar n_N} + \frac{n_{\LLL^i} }{\bar{n}_{L^i}} + \frac{n_{\HHH}}{\bar{n}_{H}}\Big)  - \sum_i\frac{\gamma_i}{2}\Big(\frac{n_{\LLL^i}}{\bar{n}_{L^i}} + \frac{{n}_{\HHH}}{ \bar n_{H}}\Big)\nonumber\\
&+& \sum_{ij}\gamma_{\mathrm{E}^{ij}}\,  \Big( \frac{n_{\EEE^j}}{\bar{n}_{E^j}} + \frac{n_{\LLL^i}}{\bar{n}_{L^i} }- \frac{n_{\HHH}}{\bar{n}_H}\Big)\,\nonumber\\
&-& \sum_{ij}\gamma_{\mathrm{U}^{ij}}\, \Big(\frac{n_{\UUU^j} }{\bar{n}_{U^j}} + \frac{n_{\QQQ^i}}{\bar{n}_{Q^i}} + \frac{n_{\HHH}}{\bar n_H}\Big) \nonumber\\
&+& \sum_{ij}\gamma_{\mathrm{D}^{ij}}\, \Big ( \frac{n_{\DDD^j}}{\bar{n}_{D^j}}  + \frac{n_{\QQQ^i}}{\bar{n}_{Q^i}}  - \frac{n_{\HHH}}{\bar n_H}\Big)\,,\nonumber\\
\dot{n}_{\Delta S}  +3H n_{\Delta S} - D_S \nabla^2 n_{\Delta S} &=& - \gamma_S  \Big( 2 \frac{n_{\NNN}}{\bar n_N} + \frac{{n}_{\Delta S}}{ \bar n_{S}}\Big)
	\per
\end{eqnarray}
In writing the $N$-number source term, we have defined $S_{\NNN}=S_N -S_{\bar{N}}$.  
These equations are valid for general flavor structures.  
The densities are defined with an implicit sum over isospin and color gauge indices.  
The multiplicity factors, which appear in \eref{eq:na_def}, are $g_N=1$, $g_{L^i}=2$, and $g_H=2$.  

\subsubsection*{Approximation in the limit of flavor universal couplings}
In the main text we analyzed a simplified limit, where flavor mixing in the right-handed neutrino is negligible.  
Moreover, for simplicity, we assume a flavor universal coupling of the lightest right-handed neutrino such that $\gamma_0^i=\gamma_0 \delta_{ii}$ and $\gamma_i=\gamma \delta_{ii}$.
Further simplifications of the above equations arise considering the size of the SM Yukawa interactions.  
The SM rates $\gamma_{\mathrm{E, U, D}}\approx \lambda_{E,U,D}^2 T^4$ do not depend on the $z$ coordinate.  
Moreover, only the third generations are in equilibrium at the temperature that we want to consider $T\approx 10^{10}\, \GeV$.  
We can therefore drop the first two generations of right-handed leptons from the diffusion equations as well as the first two generations of quarks.  
In order to get $O(1)$ estimates it is also convenient to work in the limit where the rates for third generations fermions are much faster than diffusion time scale.  
In this limit we can also drop the third generations quarks and right-handed $\tau$ lepton, since they will simply impose a constraint on the chemical potentials.

These assumptions simplify the Boltzmann equations significantly.  We define the total $L$-asymmetry as the sum of the individual asymmetries $n_{\LLL}=\sum_i n_{\LLL^i}$ (this is the quantity that appears in section \ref{sub:Diffusion} where we have dropped the $\Delta$ to simplify the notation). Given the degeneracy, we can now define $\bar{n}_L$ as the equilibrium number density for the three families, where now $g_L=6$. By these simplifications the boltzmann equations can be reduced to
\begin{eqnarray}
\dot{n}_{\NNN} + 3 H n_{\NNN} - D_N \nabla^2 n_{\NNN} &=& - \sum_i \gamma_0^i\,\Big( \frac{n_{\NNN}}{\bar n_N} + \frac{n_{\LLL} }{\bar{n}_{L}} + \frac{n_{\HHH}}{\bar{n}_{H}}\Big) - \gamma_{\Delta N}\frac{n_{\NNN}}{\bar{n}_N} +S_{\NNN}\,\nonumber\\
\dot{n}_{\LLL} + 3 H n_{\LLL}- D_L \nabla^2 n_{\LLL} &=& - \sum_i \gamma_0^i\,\Big( \frac{n_{\NNN}}{\bar n_N} + \frac{n_{\LLL} }{\bar{n}_{L}} + \frac{n_{\HHH}}{\bar{n}_{H}}\Big)- \sum_i\frac{\gamma_i}{2}\Big(\frac{n_{\LLL}}{\bar{n}_{L}} + \frac{{n}_{\HHH}}{ \bar n_{H}}\Big) \,,\nonumber\\
\dot{n}_{\HHH} +3 H n_{\HHH}- D_H \nabla^2 n_{\HHH} &=&  - \sum_i \gamma_0^i\,\Big( \frac{n_{\NNN}}{\bar n_N} + \frac{n_{\LLL} }{\bar{n}_{L}} + \frac{n_{\HHH}}{\bar{n}_{H}}\Big)- \sum_i\frac{\gamma_i}{2}\Big(\frac{n_{\LLL}}{\bar{n}_{L}} + \frac{{n}_{\HHH}}{ \bar n_{H}}\Big)\,,\nonumber\\
\end{eqnarray}
where we also neglected $\gamma$ as compared to $\gamma_{\Delta N}$ in the equation for $n_{\NNN}$. As a cross-check we can show that in the limit where diffusion does not play any role (e.g. after the completion of the phase transition, when $\gamma_0=0$), the amount of lepton asymmetry is simply controlled by
\be
\dot{n}_{\LLL} +3H n_{\LLL}=  - \frac{\sum_i\gamma^i/2}{\bar{n}_L}(n_{\LLL} +\frac{\bar{n}_{L}}{\bar{n}_{H}} n_{\HHH}).
\ee 
It is therefore important that $(\sum_i\gamma^i)/(2\bar{n}_L)$ is smaller than the Hubble rate.

\subsection{Calculation of the rates}
In this section we derive the thermally averaged rates, which are used in the main text of the paper (see \rref{Giudice:2003jh} for a review).  
We focus on the wash-out rate $\Gamma_{\wo}$ and the $N$-to-$L$ redistribution rate $\Gamma_{\LHN}$.  
By matching the full Boltzmann equations for the chemical potentials to the simplified limits discussed in the text, it is evident that the defining relations are given by
\be
	\Gamma_{\wo}=\frac{\sum_i\gamma^i/2}{\bar{n}_L}
	\qquad \text{and} \qquad 
	\Gamma_{\LHN}=\frac{\sum_i\gamma_0^i}{\bar{n}_L} 
	\per
\ee
We will work in the limit of flavor universality.  

\paragraph{Computation of $\Gamma_{\wo}$.}
Consider the process $N \to \bar{L}_i \bar{H}$.  
The corresponding transport coefficient $\gamma^i/2$ is inferred from \eref{eq:gamma_def} to be 
\be
	\frac{\gamma^i}{2} = \sum_{\rm isospin} \sum_{s_N, \, s_L} \int \! \ud \Pi_N \ud \Pi_L \ud \Pi_H \, \bar{f}_N \, | \mathcal{A}_{N \to \bar{L}_i \bar{H}}|^2 (2\pi)^4 \delta(p_N - p_L - p_H)
\ee
where we sum the $2$ final states related by isospin, and we sum the spins of $N$ and $\bar{L}_i$.  
The integral over ${\bm p}_N$ factorizes from the integrals over ${\bm p}_L$ and ${\bm p}_H$, and we can write 
\be
	\frac{\gamma^i}{2} = \bar{n}_{N,{\rm Maj}} \, \frac{K_1(m_N/T)}{K_2(m_N/T)} \, \Gamma_{N\to \bar L_i \bar H}
\ee
where $\bar{n}_{N,{\rm Maj}}$ is given by \eref{eq:nbar_nonrel} with $g_N = 2$, and 
\begin{align}
	\Gamma_{N \to \bar{L}_i \bar{H}} 
	= \frac{1}{2m_N(T)} \sum_{\rm isospin} \frac{1}{g_N} \sum_{s_N, \, s_L} \int \! \ud \Pi_L \ud \Pi_H \, | \mathcal{A}_{N \to \bar{L}_i \bar{H}}|^2 (2\pi)^4 \delta(\Sigma p)
	= 2 \times \frac{1}{2} \times \frac{\lambda_N^2}{16\pi} m_N(T)
\end{align}
is the partial width of $N$ (averaged over the initial and summed over the final spin states).  
The factor of $K_1(x)$ arises from the integration of $m_N/E_N$ with the Boltzmann factor. 
We have used $m_N \gg m_L, m_H$, since the thermal masses for $L$ and $H$ are negligible.  
The thermally-averaged washout rate (per particle), which appears in \eqref{eq:nlep_eqn}, is given by
\be\label{eq:Gamma_wo}
	\Gamma_{\wo} 
	= \frac{\sum_i\gamma^i/2}{\bar{n}_L} 
	= \frac{\lambda_N^2}{24\pi \zeta(3)} \frac{m_N(T)^3}{T^2} K_1\bigl( m_N(T) / T \bigr)
\ee
where $\bar{n}_{L} = 3 \times 2 \times (3 \zeta(3) / 4 \pi^2) T^3$ is the number density of left-chiral SM leptons, summed over generations ($3$) and isospin ($2$).  
In the regime $m_N/T \gg 1$, the washout rate is Boltzmann suppressed, $\Gamma_{\wo} \sim \lambda_N^2 m_N (m_N/T)^{3/2} e^{-m_N/T}$.  

\paragraph{Computation of $\Gamma_{\LHN}$.}
We evaluate $\gamma_0^i$ as the thermally-averaged decay rate for the process $H \to \bar{L}_i \bar{N}$.  
Outside of the bubbles, both the electroweak and $\U{1}_{\rm L}$ symmetries are unbroken, and the particle masses arise entirely from thermal effects.  
For the parameters of interest we have $m_H > m_L + m_N$, and therefore the Higgs decay channel is open.  
(If the 1-to-2 processes are kinematically blocked or suppressed, the 2-to-2 scattering $WH \to \bar{L}_i \bar{N}$ will mediate the transfer of $N$-number into $L$-number.  If we were to include these processes mediated by a $t$-channel fermion exchange, $\gamma_0^i$ is enhanced by an $O(1)$ factor \cite{Garbrecht:2013urw}.)  
We calculate $\gamma_0^i$ in the same way as $\gamma^i$ above, and we find 
\be
	\gamma_0^i = 2 \times \frac{m_H^2 T}{2\pi^2} K_1(m_H/T) \, \Gamma_{H \to \bar L_i \bar N}
\ee
where the factor of $2$ accounts for a sum over isospin.  
The partial width of $H$ is given by $\Gamma_{H \to \bar{L}_i\bar{N}} = \bigl( \lambda_N^2m_H(T) \bigr) / (16\pi)$, and there is no sum on the final state spins because $\bar{L}_i$ and $\bar{N}$ are chiral.  
From the Boltzmann equation we now get the definition of $\Gamma_{\LHN}$,
\be\label{eq:Gamma_LHN}
	\Gamma_{\LHN} 
	= \frac{\sum_i \gamma_0^i}{\bar{n}_L} 
	= \frac{\lambda_N^2}{24\pi \zeta(3)} \frac{m_H(T)^3}{T^2} K_1\bigl( m_H(T) / T \bigr)
\ee
where $\bar{n}_{L}$ appears below \eref{eq:Gamma_wo}.  
In the regime $m_H(T) \ll T$, we have $(m_H^3/T^2) K_1(m_H/T) \to (m_H^2/T)$.  

\end{appendix}

\bibliographystyle{JHEP}
\bibliography{refs--composite_leptogen}

\end{document}